\documentclass[twocolumn,showpacs,preprintnumbers,amsmath,amssymb]{revtex4-1}

\usepackage{graphicx}
\usepackage{dcolumn}
\usepackage{bm}
\usepackage{epstopdf}
\usepackage{lineno}
\usepackage[caption=false]{subfig} 

\begin{document}


\title{Panic contagion and the evacuation dynamics} 

\author{F.E.~Cornes}
 \affiliation{Departamento de F\'\i sica, Facultad de Ciencias 
Exactas y Naturales, Universidad de Buenos Aires,\\
 Pabell\'on I, Ciudad Universitaria, 1428 Buenos Aires, Argentina.}
 \author{G.A.~Frank}
 \affiliation{Unidad de Investigaci\'on y Desarrollo de las 
Ingenier\'\i as, Universidad Tecnol\'ogica Nacional, Facultad Regional Buenos 
Aires, Av. Medrano 951, 1179 Buenos Aires, Argentina.}
\author{C.O.~Dorso}%
 \email{codorso@df.uba.ar}
\affiliation{Instituto de F\'\i sica de Buenos Aires}
 \affiliation{Departamento de F\'\i sica, Facultad de Ciencias 
Exactas y Naturales, Universidad de Buenos Aires,\\
 Pabell\'on I, Ciudad Universitaria, 1428 Buenos Aires, Argentina.}

\date{\today}

\begin{abstract}
Panic may spread over a crowd in a similar fashion as contagious diseases do in 
social groups. People no exposed to a panic source may express 
fear, alerting others of imminent danger. This social mechanism initiates an 
evacuation process, while affecting the way people try to escape.  We 
examined real life situations of panic contagion and reproduced these 
situations in the context of the Social Force Model. We arrived to the 
conclusion that two evacuation schemes may appear, according to the 
\textit{stress} of the panic contagion. Both schemes exhibit different 
evacuation patterns and are qualitatively visible in the available real life 
recordings of crowded events. We were able to quantify these patterns through
topological parameters. We further investigated how the panic spreading 
gradually stops if the source of danger ceases.
\end{abstract}


\maketitle

\section{\label{introduction}Introduction}

Many authors called the attention on the fact that panic is a contagious 
phenomenon \cite{volchenkov,zhao,fu2014,fu2016}. Panic may spread over any 
simple ``social group'' if some kind of coupling mechanism exists between 
agents \cite{volchenkov}. This coupling mechanism corresponds to social 
communication appearing in the group. As a consequence, the individuals 
(agents) may change their anxiety state from relaxed to a panic one (and back 
again) \cite{volchenkov}.   \\

Panic contagion over the crowd can be attained if the coupling mechanism 
between individuals is strong enough and affects many neighboring pedestrians 
\cite{volchenkov}. Research on random lattices shows that the coupling 
stress becomes relevant whenever the number on neighbors is small 
(\textit{i.e.} less than four). That is, a small connectivity number between 
agents (pedestrians) requires really moving gestures \cite{volchenkov}.  \\

Recent investigation suggests that other psychological mechanisms than social
communication can play an important role during the panic spreading 
over the crowd \cite{zhao,fu2014,chen2015}. Susceptibility appear as relevant 
attributes that control the panic propagation \cite{zhao}. Consequently, 
diseases contagion models are usually introduced when studying the panic 
spreading. The Susceptible-Infected-Recovered-Susceptible (SIRS) model raises 
as a suitable research tool for examining the panic dynamics. 
The spreading model is, therefore, represented as a system of first order 
equations \cite{zhao,chen2015}. \\


According to the SIRS model implemented in Ref.~\cite{zhao}, a dramatic 
contagion of panic can be expected in those crowded situations where the 
individuals are not able to calm down quickly. The speed at which the 
individual calms down may not only depend on the current environment, but on 
other psychological attributes \cite{zhao}. Ref.~\cite{Ta} proposes a 
characteristic value for this ``stress decay''.  \\

Although the SIRS model appears to be a reasonable approach to panic spreading, 
it has been argued that it may not accurately resemble the situations of crowds 
with moving pedestrians \cite{fu2014,fu2016}. The moving pedestrians will 
get into panic if their ``inner stress'' exceeds a threshold \cite{fu2016}. 
That is, if the cumulative emotions received by the pedestrian's neighbors 
surpasses a certain ``inner stress'' threshold. \\

Conversely, unlike the SIRS model, any panicking pedestrian may relax after 
some time due to ``stress decay'' (if no emotions of fear are received by the 
corresponding neighbors) \cite{fu2014,fu2016}. That is, in this case, there is 
not a probability to switch from the anxious (infected) state to the relaxed 
(recovered) state as in the SIRS model, but a natural decay. Thus, the 
increase in the ``inner stress'' and the ``stress decay'' are actually the two 
main phenomena attaining for the pedestrians behavior. \\

Researchers seem not to agree on how the increase in the ``inner stress'' and 
the ``stress decay'' affect the pedestrians behavioral patterns 
\cite{pelechano,fu2016,nicolas}. Pelechano and co-workers \cite{pelechano} 
suggest that the maximum \textit{current} velocity of the pedestrians may 
increase if he (she) gets into panic. But Fu and co-workers \cite{fu2016} 
propose to update the \textit{desired} velocity (not the current one) of the 
pedestrian, according to his (her) current ``inner stress'' (see Section 
\ref{background} for details). Both investigations assume that the pedestrians 
move in the context of the Social Force Model (SFM).  \\  

More experimental data needs to be examined before arriving to consensus on how 
the panic contagion affects the pedestrians dynamics. \\

Our investigation focuses on two real life situations. Our aim is to develop a 
model for describing striking situations, where many individuals may suddenly 
switch to an anxious state. We will focus on video analyses in order to obtain 
reliable parameters from a real panic-contagion events, and further test these 
parameters on computing simulations. \\


In Section \ref{background} we introduce the dynamic equations for evacuating 
pedestrians, in the context of the Social Force Model (SFM). We also define 
the meaning of the appearance to danger, the \textit{contagion stress} 
and their relation to the pedestrians desired velocity.  \\

In Section \ref{experimental} we present the real life situations considered 
in our investigation. The corresponding simulations (in the SFM context) 
resembling these situations are detailed in Section \ref{simulations}. \\

Section \ref{results} exhibits the results of our investigations, while some 
preliminary outcomes are summarized. Section \ref{conclusions} details the 
corresponding conclusions.   \\

\section{\label{background}Background}

\subsection{\label{sfm}The social force model}

This investigation handles the pedestrians dynamics in the context of the 
``social force model'' (SFM) \cite{Helbing1}. The SFM exploits the idea 
that human motion depends on the people's own desire to reach a certain 
destination, as well as other environmental factors \cite{Helbing4}. The former 
is modeled by a force called the ``desire force'', while the latter is 
represented by social forces and ``granular forces''. These forces enter the 
motion equation as follows

\begin{equation}
m_i\,\displaystyle\frac{d\mathbf{v}^{(i)}}{dt}=\mathbf{f}_d^{(i)}
+\displaystyle\sum_{j=1}^{N}\displaystyle\mathbf{f}_s^{(ij)}
+\displaystyle\sum_ {
j=1}^{N}\mathbf{f}_g^{(ij)}\label{eq_mov}
\end{equation}

\noindent where the $i,j$ subscripts correspond to any two pedestrians in the 
crowd. $\mathbf{v}^{(i)}(t)$ means the current velocity of the pedestrian  
$(i)$, while $\mathbf{f}_d$ and $\mathbf{f}_s$ correspond to the ``desired 
force'' and the ``social force'', respectively. $\mathbf{f}_g$ is the friction 
or granular force. \\

The $\mathbf{f}_d$ attains the pedestrians own desire to reach  
a specific target position at the desired velocity $v_d$. But, due to 
environmental factors (\textit{i.e.} obstacles, visibility), he (she) actually 
moves at the current velocity $\mathbf{v}^{(i)}(t)$. Thus, the acceleration (or 
deceleration) required to reach the desired velocity $v_d$ corresponds to the 
aforementioned ``desire force'' as follows  
\cite{Helbing1}

\begin{equation}
        \mathbf{f}_d^ {(i)}(t) =  
m_i\,\displaystyle\frac{v_d^{(i)}\,\mathbf{e}_d^
{(i)}(t)-\mathbf{v}^{(i)}(t)}{\tau} \label{desired}
\end{equation}

\noindent where $m_i$ is the mass of the pedestrian $i$ and $\tau$ represents 
the relaxation time needed to reach the desired velocity. $\mathbf{e}_d$ 
is the unit vector pointing to the target position. Detailed values for $m_i$ 
and $\tau$ can be found in Refs.~\cite{Helbing1,Dorso3}.\\

Besides, the ``social force'' $\mathbf{f}_s(t)$ represents the 
socio-psychological tendency of the pedestrians to preserve their \emph{private 
sphere}. The spatial preservation means that a repulsive feeling exists between 
two neighboring pedestrians, or, between the pedestrian and the walls 
\cite{Helbing1,Helbing4}. This repulsive feeling becomes stronger as people get 
closer to each other (or to the walls). Thus, in the context of the social 
force 
model, this tendency is expressed as 

\begin{equation}
        \mathbf{f}_s^{(ij)} = A_i\,e^{(r_{ij}-d_{ij})/B_i}\mathbf{n}_{ij} 
        \label{social}
\end{equation}

\noindent where $(ij)$ corresponds to any two pedestrians, or to the 
pedestrian-wall interaction. $A_i$ and $B_i$ are two fixed parameters (see 
Ref.~\cite{Dorso1}). The distance $r_{ij}=r_i+r_j$ is the sum of the 
pedestrians radius, while $d_{ij}$ is the distance between the center of mass 
of the pedestrians $i$ and $j$. $\mathbf{n}_{ij}$ means the unit vector in the 
$\vec{ji}$ direction. For the case of repulsive feelings with the walls, 
$d_{ij}$ corresponds to the shortest distance between the pedestrian and the 
wall, while $r_{ij}=r_i$ \cite{Helbing1,Helbing4}.  \\

It is worth mentioning that the Eq.~(\ref{social}) is also valid if two 
pedestrians are in contact (\textit{i.e.} $r_{ij}>d_{ij}$), but its meaning is 
somehow different. In this case, $\mathbf{f}_s$ represents a body repulsion, as 
explained in Ref.~\cite{Dorso5}.\\

The granular force $\mathbf{f}_g$ included in Eq.~(\ref{eq_mov}) corresponds 
to the sliding friction between pedestrians in contact, or, between pedestrians 
in contact with the walls. The expression for this force is 

\begin{equation}
        \mathbf{f}_g^{(ij)} = 
\kappa\,(r_{ij}-d_{ij})\,\Theta(r_{ij}-d_{ij})\,\Delta
\mathbf{v}^{(ij)}\cdot\mathbf{t}_{ij} 
        \label{granular}
\end{equation}

\noindent where $\kappa$ is a fixed parameter. The function 
$\Theta(r_{ij}-d_{ij})$ is zero when its argument is negative (that is, 
$r_{ij}<d_{ij}$) and equals unity for any other case (Heaviside function). 
$\Delta\mathbf{v}^{(ij)}\cdot\mathbf{t}_{ij}$ represents the difference between 
the tangential velocities of the sliding bodies (or between the individual and 
the walls).   \\

\subsection{\label{inner}The inner stress model}

As mentioned in Section \ref{introduction}, the ``inner stress'' stands for 
the cumulative emotions that the pedestrian receives from his (her) neighbors. 
This magnitude may change the pedestrian's behavior from a relaxed state to 
panic, and consequently, we propose that his (her) desired velocity $v_d$ 
increases as follows \cite{fu2016}

\begin{equation}
 v_d(t)=[1-M(t)]\,v_d^\mathrm{min}+M(t)\,v_d^\mathrm{max}\label{inner_strength}
\end{equation}

\noindent for $M(t)$ representing the ``inner stress'' as a function of time. 
The minimum desired velocity $v_d^\mathrm{min}$ corresponds to the (completely) 
relaxed state, while the maximum desired velocity $v_d^\mathrm{max}$ 
corresponds to the (completely) panic state.  \\

The inner stress $M(t)$ in Eq.~(\ref{inner_strength}) is assumed to be 
bounded between zero and unity. Vanishing values of $M(t)$ mean that the 
pedestrian is relaxed, while values approaching unity correspond to a very 
anxious pedestrian (\textit{i.e.} panic state). \\

The emotions received from the pedestrian's surrounding are responsible for the 
increase in his (her) inner stress $M(t)$. But, in the absence of stressful 
situations, some kind of relaxation occurs (say, the ``stress decay''), 
attaining a decrease in $M(t)$. Following Ref.~\cite{nicolas}, a first order 
differential equation for the time evolution of $M(t)$ can be assumed

\begin{equation}
\displaystyle\frac{dM}{dt}=-\displaystyle\frac{M}{\tau_M}
+ \mathcal { P } \label{PM}
\end{equation}

The differential ratio on the left of Eq.~(\ref{PM}) expresses the change in 
the ``inner stress'' with respect to time. Whenever the pedestrian receives 
alerting emotions from his (her) neighbors (expresses by the contagion 
efficiency $\mathcal{P}$), the ``inner stress'' is expected to increase. But, 
if no alerting emotions are received, his (her) stress is expected to decay 
according to a fixed relaxation  time $\tau_M$. Thus, the first term on the 
right of Eq.~(\ref{PM}) handles the settle down process towards the relaxed 
state. The second term on the right, on the contrary, increases his (her) 
stress towards an anxious state. \\  

\begin{figure}
\includegraphics[width=0.8\columnwidth]{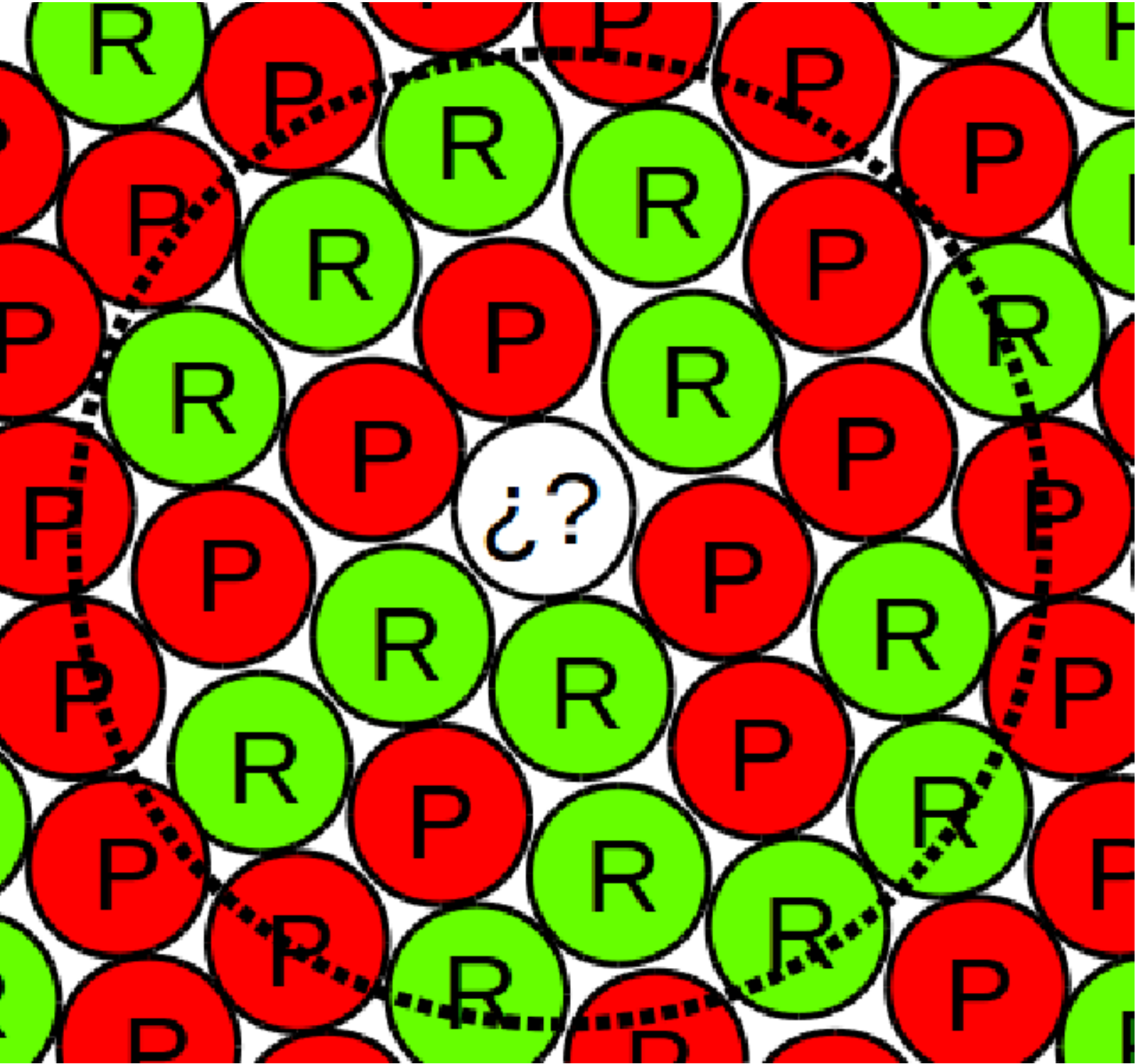}
\caption{\label{panic_diagram} (Color on-line only) Crowd with relaxed (green 
circles) and panicking (red circles) pedestrians. The state of the pedestrian 
labeled with question marks depends of the amount of neighboring panicking and 
relaxed individuals (see Eq.~\ref{eqn_contagion_prob}). The circle indicate the 
 contagious radius.}
\end{figure}

We assume that the parameter $\mathcal{P}$ attains the emotions received from 
alerting (anxious) neighbors within a  certain radius, called the 
\textit{contagion radius}. As described in Appendix \ref{appendix_1}, if $k$ 
pedestrians among $n$ neighbors are expressing fear (see 
Fig.~\ref{panic_diagram}), then the actual value of $\mathcal{P}$  is

\begin{equation}
\mathcal{P}=J\,\bigg\langle\displaystyle\frac{k}{n}\bigg\rangle
\label{eqn_contagion_prob}
\end{equation}

\noindent where the parameter $J$ represents an \textit{effective contagion 
stress} (see Appendix \ref{appendix_1} for details). This parameter 
resembles the pedestrian susceptibility to enter in panic. For simplicity we 
further assume that this parameter is the same for all the pedestrians.  \\

The symbol $\langle\cdot\rangle$ represents the mean value for any short 
time interval (see Appendix \ref{appendix_1} for details). However, for 
practical reasons, we will replace this mean value with the sample value $k/n$ 
at each time-step. \\

\subsection{\label{contagion}The stress decay model}

The pedestrian ``stress decay'' corresponds to the individual's natural 
relaxation process in the absence of stimuli (\textit{i.e.} emotions), until he 
(she) settles to relaxed. This behavior is mathematically expressed through 
the relaxation term in Eq.~(\ref{PM}). Thus, in the absence of stimuli (that 
is, vanishing values of $\mathcal{P}$), it follows from 
Eq.~(\ref{inner_strength}) and Eq.~(\ref{PM}) that  \\

\begin{equation}
v_d(t)=v_d^\mathrm{min}+(v_d^\mathrm{max}-v_d^\mathrm{min})\,M(0)\,e^{-t/\tau_M}
\label{vd_relaxation}
\end{equation}

\noindent for any fixed value $M(0)$ at $t=0$, and a vanishing value of 
$M(t)$ long time after ($t\gg\tau_M$). The characteristic time $\tau$ is 
different from $\tau_M$. Ref.~\cite{Ta} suggests that $\tau_M\simeq 50\,$ 
seconds. 
 \\  

The characteristic time $\tau_M$ may be different from the suggested value 
according to specific environmental factors. Eq.~(\ref{vd_relaxation}) 
proposes the way to handle an estimation of $\tau$ whenever the composure 
desired velocity $v_d(t_c)$ is known ($t_c$ being the time required to arrive 
to composure). Assuming $M(0)=1$, it is straight forward that

\begin{equation} 
\tau_M^{-1}=\displaystyle\frac{1}{t_c}\,\ln\bigg(\displaystyle
\frac{v_d^\mathrm{max}-v_d^\mathrm{min}}
{v_d(t_c)-v_d^\mathrm { min } }
\bigg) \label{tau}
\end{equation}

\subsection{\label{Minkowski}Topological characterization}

One of the most useful image processing technique is the computation of the 
Minkowski functionals. This general method, based on the concept of integral 
geometry, uses topological and geometrical descriptors to characterize the 
topology of two and three dimensional patterns. \\

We used this method to analyze data (images) obtained from the video of the 
Charlottesville incident. So, we focused the attention on the 2-D case. Three 
image functionals can actually be defined in 2-D: area, perimeter and 
the Euler characteristic. The three can give a complete description of 2-D 
topological patterns appearing in (pixelized) black and white images. \\

To characterize a pattern on a black and white image, each black (or white) 
pixel is decomposed into 4 edges, 4 vertices and the interior of the pixel or 
square. Taking into account the total number of squares ($n_s$), edges ($n_e$) 
and vertices ($n_v$), the area ($A$), perimeter ($U$) and Euler 
characteristic ($\chi$) are defined as   
 
\begin{equation}
A = n_s,~~U = -4n_s+2n_e,~~\chi = n_s-n_e+n_v
\label{mink}
\end{equation}

The area is simply the total number of (black or white) pixels. The second and 
third Minkowski functionals describe the boundary length and the connectivity 
or topology of the pattern, respectively. The latter corresponds to the 
number of surfaces of connected black (white) pixels minus the number of 
completely enclosed surfaces of white (black) pixels (see 
Ref.~\cite{Michielsen}). \\

\section{\label{experimental}Experimental data}

In this section we introduce two incidents, as examples of real life panic 
propagation. The first one occurred in Turin (Italy) while the other one took 
place in Charlottesville (USA) in 2017. We further present relevant data 
extracted from the corresponding videos available in the web (see on-line 
complementary material). \\

\subsection{\label{expturin} Turin (Italy)}

\begin{figure*}[!htbp]
\hfill
\subfloat[Pedestrians watching the screen.\label{fig:original_frame_turin}]{
\includegraphics[width=0.67\columnwidth]{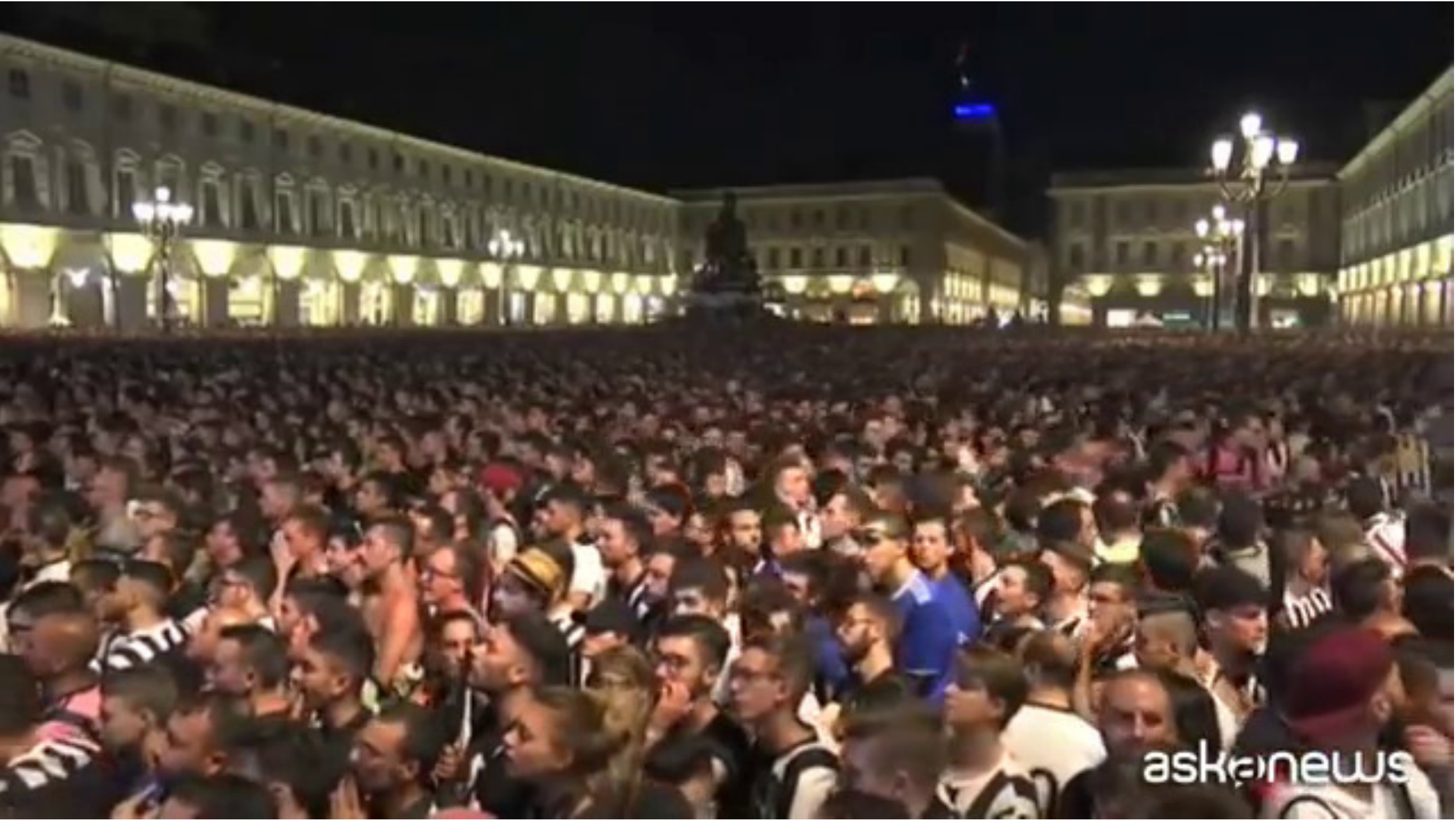}
}
\hspace{10mm}
\subfloat[Pedestrians in panic.\label{fig:media_frame_turin}]
{\includegraphics[width=0.67\columnwidth]{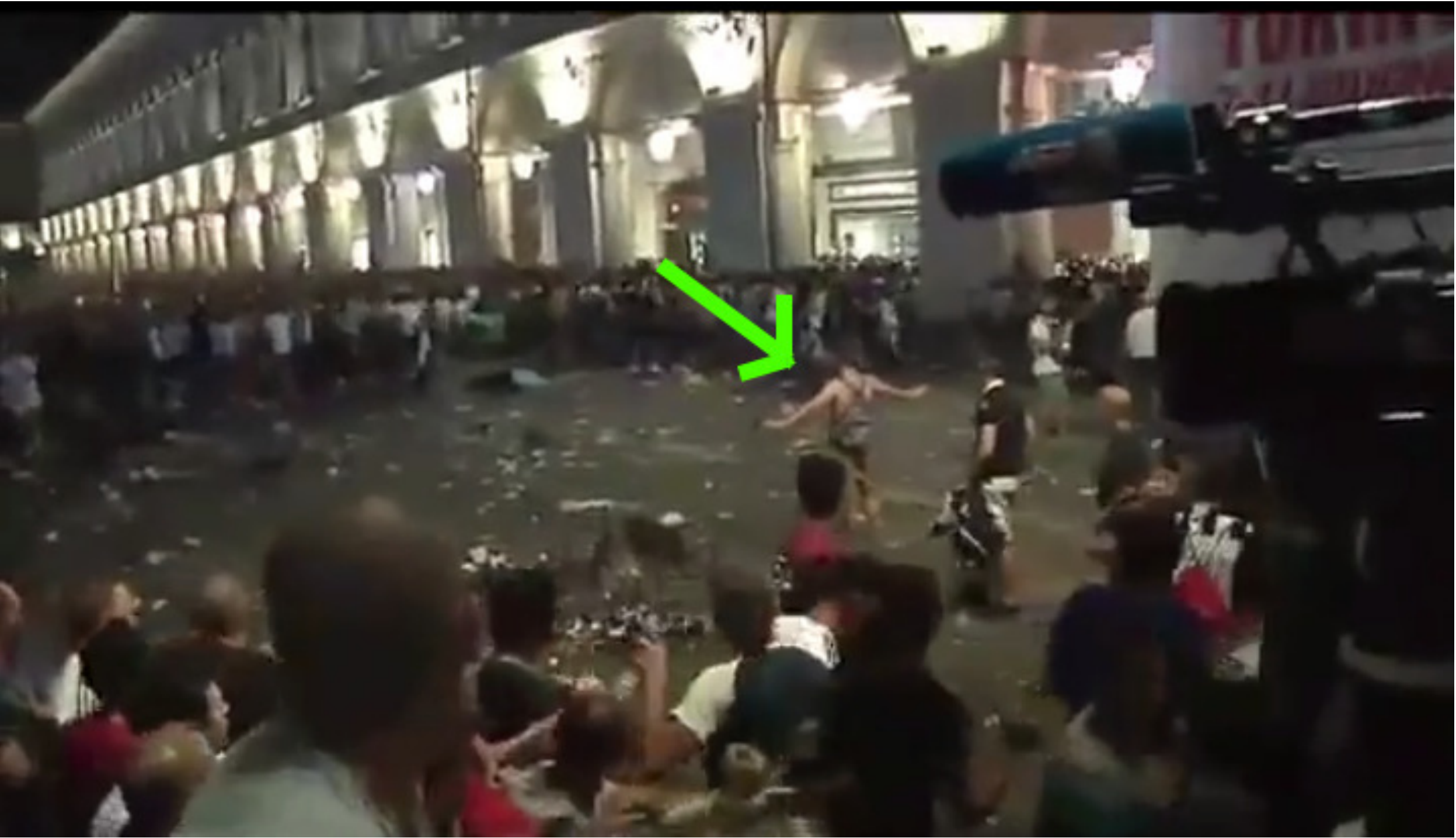}
}\hfill
\subfloat[Panic spreading.\label{fig:analysis_frame_turin}]{
\includegraphics[width=0.53\columnwidth]{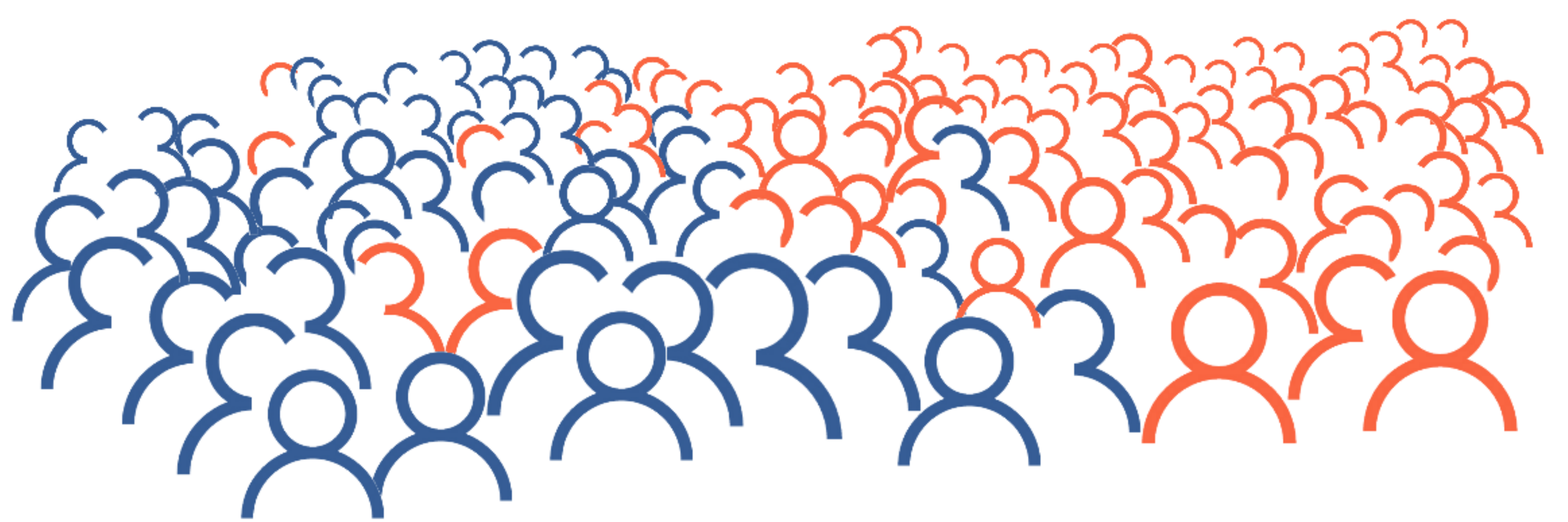}
}
\caption{\label{fig:experimentaldataturin} (Color on-line only) (a) 
Snapshot of the crowd watching the football match. The screen is on 
the left (out of the scene). The pedestrians on the right are actually in 
fear due to the \textit{fake bomber}.  The (b) snapshot corresponds to the same 
scene as (a) but shifted  to the right (actually, the camera appearing in this 
image is the one that captured the (a) image). The \textit{fake bomber} appears 
in the scene and is indicated with a green arrow. (c) Analysis of the panic 
spreading among the crowd. The blue and orange profiles represent the relaxed 
and anxious pedestrians, respectively, associated to the (a) image (see text 
for 
details). In fact, in the (b) image we can observe (on the right of the image) 
the camera that captured the (a) image. The total number of contour bodies is 
$N=131$.} 
\end{figure*}

On June 3rd 2017, many Juventus fans were watching the Champions League final 
between Juventus and Real Madrid on huge screens at Piazza San Carlo. During 
the second half of the match, a stampede occurred when one (or more) 
individuals shouted that there was a bomb. More than 1000 individuals were 
injured during the stampede, although it was a false alarm. 
Fig.~\ref{fig:experimentaldataturin} captures two moments of the panic 
spreading (see caption for details). The arrow in 
Fig.~\ref{fig:media_frame_turin} points to the individual that caused the panic 
spreading. He will be called the  \textit{fake bomber} throughout this 
investigation. \\

The recordings from Piazza San Carlo show how the pedestrians escape 
away from the ``panic source'', that is, from the \textit{fake bomber}. It can 
be seen in Fig.~\ref{fig:media_frame_turin} the opening around the panic source 
a few seconds after the shout. The opening exhibits a circular pattern around 
the \textit{fake bomber}. This pattern gradually slows down as the pedestrians 
realize the alarm being false. Approximately 20 seconds after the shout, the 
pedestrians calm down to the relaxed state while the opening closes. \\

In order to quantify the panic contagion among the crowd, we split the video 
into 14 images. The frame rate was 2 frames per second. Thus, the time 
interval between successive images was 0.5~seconds. This time interval 
corresponds to the acceleration time $\tau$ in the SFM.\\

Fig.~\ref{fig:analysis_frame_turin} shows the profile corresponding to the 
first image. Any (distinguishable) pedestrian in 
Fig.~\ref{fig:original_frame_turin} is outlined in
Fig.~\ref{fig:analysis_frame_turin} as a body contour. The contour colors 
represent relaxed pedestrians (\textit{i.e.} blue in the on-line version) 
or pedestrians in panic (\textit{i.e.} orange in the on-line version). The 
latter correspond to the individuals that suddenly changed their motion 
pattern. That is, individuals that turned back to see what happened or 
pedestrians that were pushed towards the screen (on the left) due to the 
movement of his (her) neighbors. \\

The panic spreading shown in Fig.~\ref{fig:analysis_frame_turin} occurs from 
right to left, until nearly all the contour bodies switch to the panic state 
(\textit{i.e.} orange in the on-line version). Notice, however, that a few 
pedestrians may remain relaxed for a while, even though his (her) neighbors 
have already switched to the panic state. Or, on the contrary, pedestrians in 
panic may be completely surrounded by relaxed pedestrians, as appearing on the 
left of Fig.~\ref{fig:analysis_frame_turin}. Both instances are in agreement 
with the hypothesis that pedestrians may switch to a panic state according to 
an \textit{contagion efficiency} $\mathcal{P}$. See Appendix \ref{appendix_1} 
for details on the $\mathcal{P}$ computation within the contagion radius.  \\

The inspection of successive images provides information on the new anxious or 
panicking pedestrians and the state of their current neighbors. Appendix 
\ref{appendix_2} summarizes this information, while detailed values for the 
contagion efficiency $\mathcal{P}$ and the contagion stress $J$ are reported 
in Table~\ref{table:1}. Notice that the data sampling is strongly limited by 
the total number of outlined pedestrians (that is, 131 individuals). Thus, the 
reported values for $t>4\,$s are not really suitable as parameter estimates 
because of the finite size effects. In order to minimize the size effects, we 
focused on the early stage of the contagion were the contagion stress $J$ 
seems to be (almost) stationary (see Fig.~\ref{intensity}). \\

The (mean) contagion stress for the Turin incident was found to 
be $J=0.1\pm0.055$ (within the standard deviation). This value appears to be 
surprisingly low according to explored values in the literature (see 
Ref.~\cite{nicolas}). However, we shall see in Section \ref{results} that this 
stress is enough to reproduce real life incidents.  \\

\subsection{\label{expvirginia} Charlottesville, Virginia (USA) }

One person was killed and 19 injured when a car ran over into a crowd of  
pedestrians during an antifascist protest (Charlottesville, August 
12th, 2017). The incident took place at the crossing of Fourth St. and Water 
St. Fig.~\ref{fig:original_frame} shows a snapshot of the incident (the video 
is provided in the supplementary material).  \\

\begin{figure*}[!htbp]
\subfloat[Original frame.\label{fig:original_frame}]{
\includegraphics[width=0.70\columnwidth]
{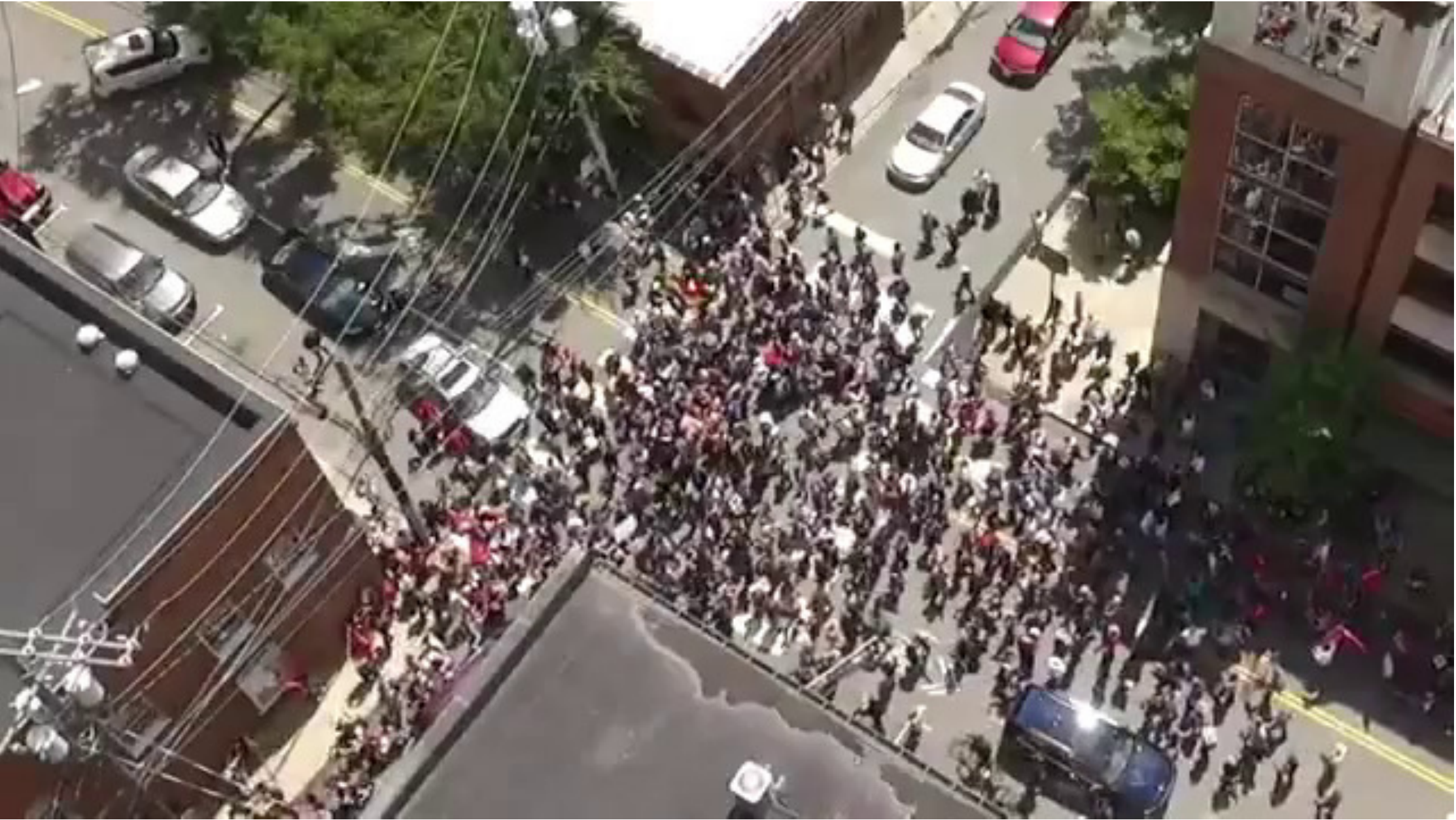}}
\subfloat[Pixelated image. \label{fig:pixel_frame}]{
\includegraphics[width=0.70\columnwidth]
{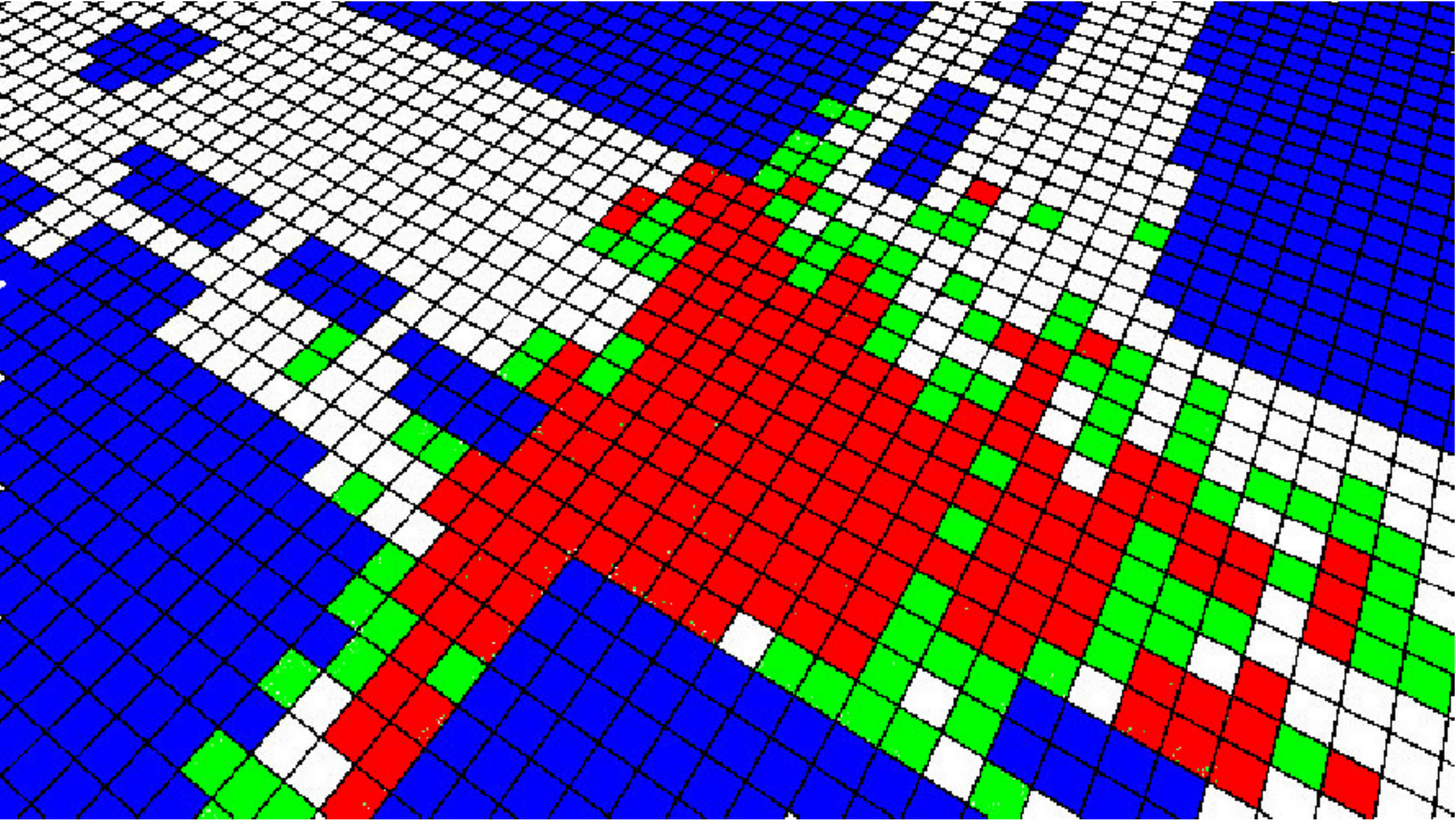}
}
\subfloat[Perspective correction. \label{fig:final_frame}]{
\includegraphics[width=0.60\columnwidth]{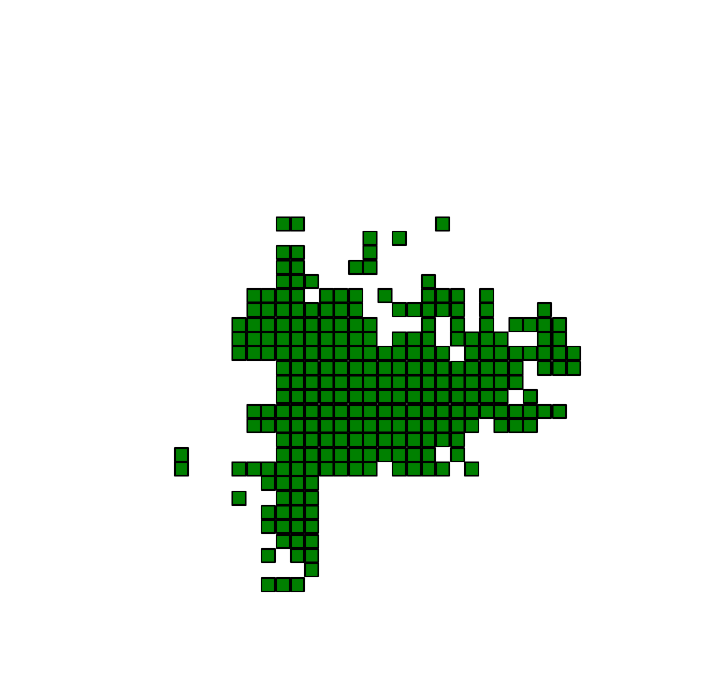}
}
\caption{\label{fig:experimentaldataCharlottesville} (a) Image of the incident 
in Charlottesville. This image corresponds to the first frame of the 
video. (b) Pixelated image of the original frame. In the image we identify in 
blue color the obstacles, like cars and buildings. Green and red cells are 
occupy by one and more than one pedestrians, respectively. The street was 
colored in white. The line spacing was 12 pixels. The cell size was, 
approximately, $1.5\,\mathrm{m}\times1.5\,\mathrm{m}$. (c) Perspective 
correction of the pixelated image. In green color we represents the position of 
the occupied cells by pedestrians. The white spaces represents the obstacles 
(buildings and cars) and the street.} 
\end{figure*}

In the video, we can see that the whole crowd gets into panic. But, we can 
identify two groups of pedestrians, according to the amount of information 
they have about the incident. The individuals near the car (say, less than 5~m) 
actually witnessed when the driver ran over into the crowd. However, far 
away pedestrians become aware that something happened among the crowd due to  
the fear emotions of his (her) neighbors. But, they cannot determine the 
nature of the incident because the car is out of their sight. Thus, the 
pedestrians nearer to the car have more information than the far away 
individuals.\\

The video also shows that the pedestrians close to the car stop running as soon 
as the car stops. On the contrary, far away individuals continue escaping 
after this occurs due to their lack of information. \\

\begin{figure}
\includegraphics[width=\columnwidth]{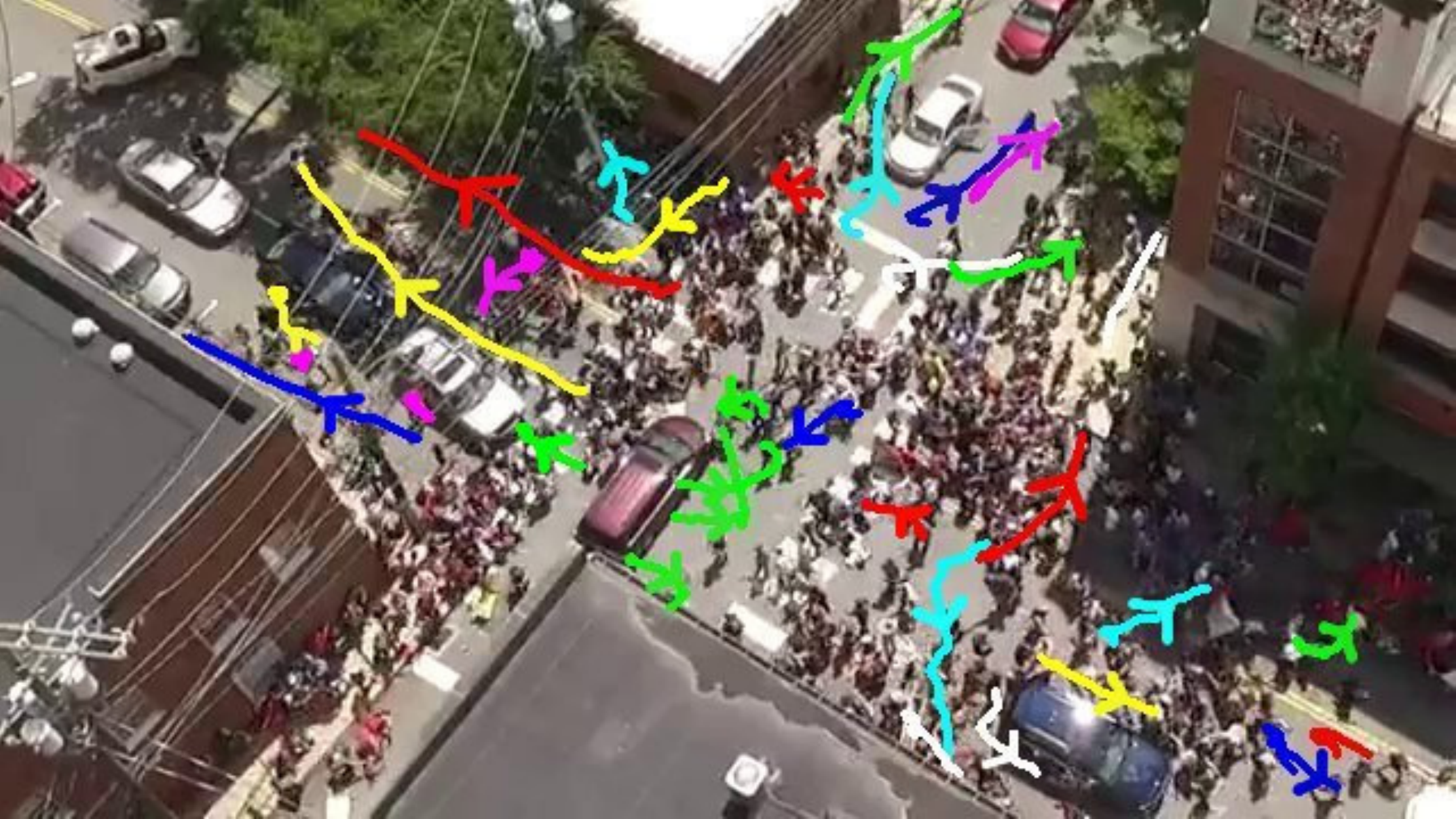}
\caption{\label{tracks} (Color on-line only) Trajectories for some 
pedestrians in panic from the Charlottesville video. The pedestrians' 
positions were recorded on consecutive images, and then joined by means of the 
software ImageJ. The arrows represent the movement direction.}
\end{figure}

Using the program ImageJ \cite{Rasband}, we were able to follow the 
trajectories of various pedestrians. As shown in Fig.~\ref{tracks}, most of the 
trajectories are approximately radial to the car. Notice, however, that three 
individuals ran toward the car to help the other injured pedestrians. \\

In order to obtain more experimental data, we split the video into 19 
frames. The frame rate was two frames per second. We further overlapped a 
square grid on each frame, but taking into account the two-point perspective of 
each image. Each cell was colored with different colors depending if it was 
occupied by pedestrians, obstacles, etc (see caption in 
Fig.~\ref{fig:pixel_frame} for details). Finally, we performed a 
back-correction of the perspective for a better inspection of the grid. The 
result is shown in Fig.~\ref{fig:final_frame}. \\

The complete analysis of the geometrical and topological patterns appearing on 
the grid can be found in Section \ref{results}.  \\

\section{\label{simulations}Numerical simulations}

\subsection{\label{boundary} The simulation conditions}

\subsubsection*{\label{sim_turin}The Turin scenario}

\begin{figure*}[!htbp]
\subfloat[Turin initial state. 
\label{fig:sim_turin}]{\includegraphics[scale=0.35]
{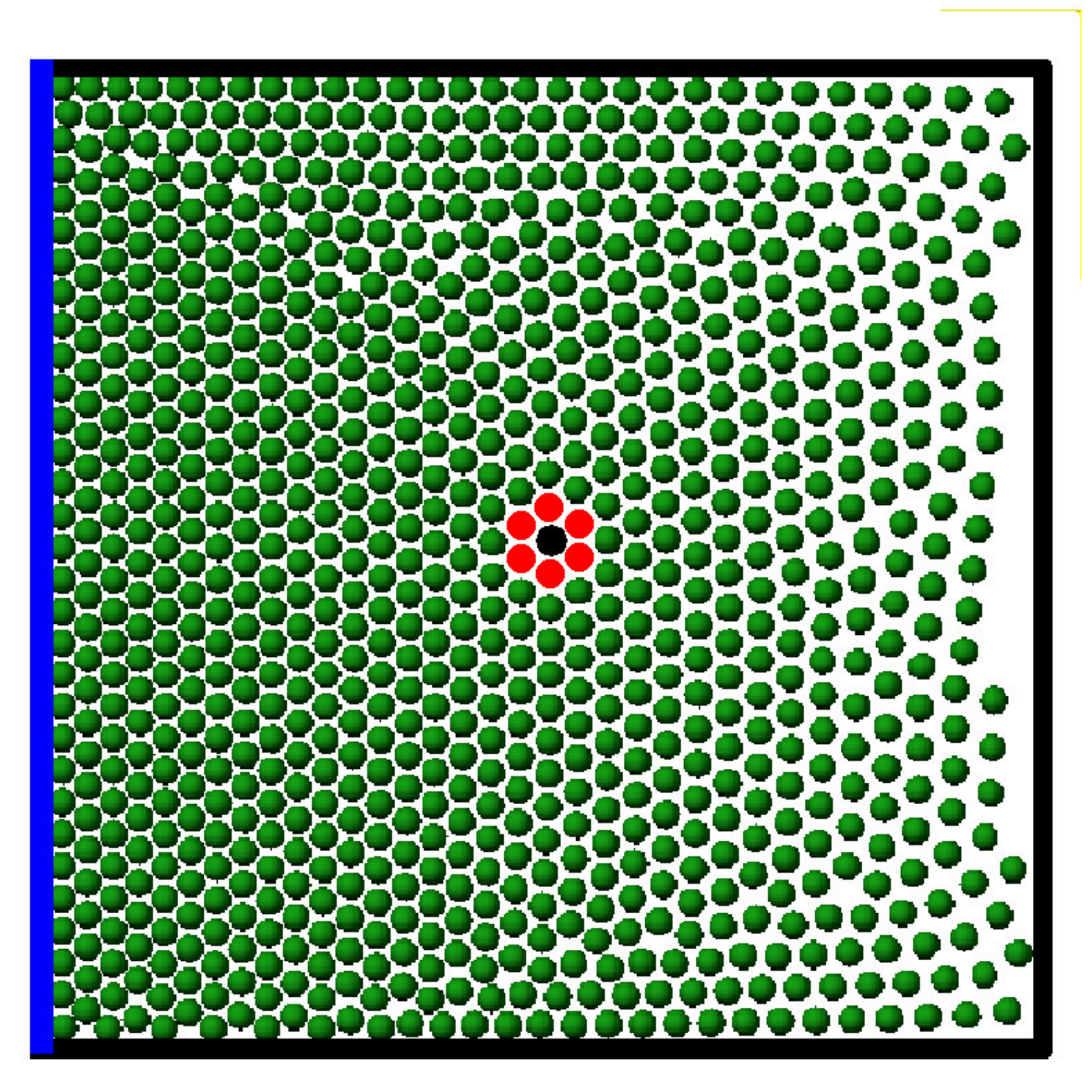}
}\hspace{25mm}
\subfloat[Charlottesville initial state. 
\label{fig:sim_virginia}]{
\includegraphics[scale=0.35]
{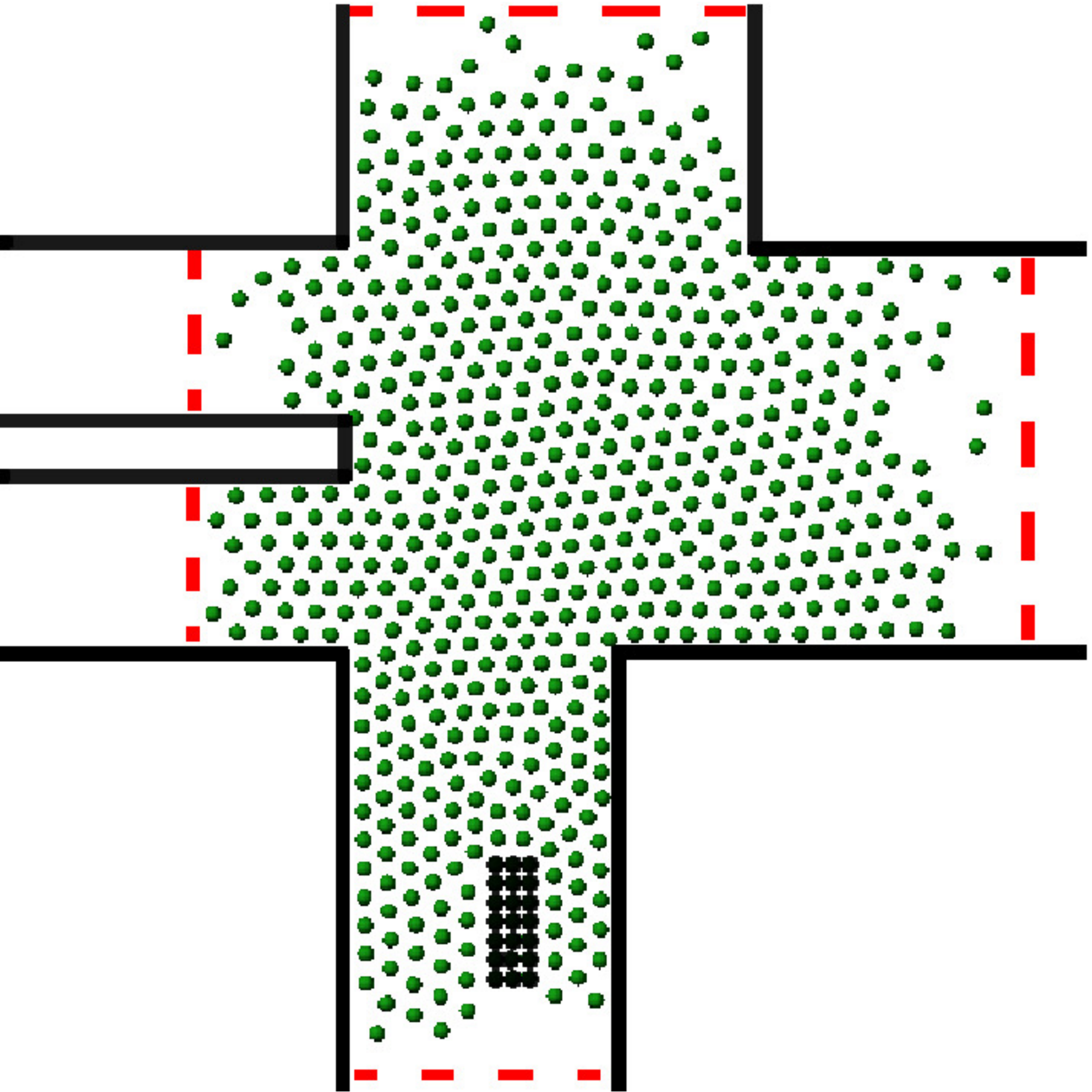}
}
\caption{\label{fig:simulations} Snapshots of the initial configuration for the 
(a) Turin (Italy) and (b) Charlottesville (Virginia, USA) simulations. The 
relaxed pedestrians are represented in green circles on both images. (a) The 
\textit{fake bomber} is represented in black, while his first neighbors are 
represented in red (see text for details). The blue line on the left 
represents the wide screen. (b) The car is represented by 21 black circles and 
moves from bottom to top. The solid (black) lines represent the walls and 
other obstacles (\textit{i.e.} parked cars on the middle left of 
the image (see Fig.~\ref{fig:experimentaldataCharlottesville})). The width of 
the street at the bottom of the image is 8~m and at the top reaches 15~m (see 
red dashed lines). The street width on the left is divided into two paths of 
5~m each, while the street width on the right is 13~m (see 
red dashed lines). Initially, the pedestrians were placed at random positions 
inside the region bounded by the red dashed lines.} 
\end{figure*}

We mimicked the Turin incident (see Section \ref{experimental}) by first 
placing 925 pedestrians inside a 21~m $\times$ 21~m square region. The 
pedestrians were placed in a regular square arrangement, meaning that the 
occupancy density was approximately 2~people/m$^2$. After their desire force 
was set (see below), the crowd was allowed to move freely until the 
pedestrian's velocity vanished. This balance situation can be seen in 
Fig.~\ref{fig:sim_turin} and corresponds to the initial configuration for the 
panic spreading simulation.  \\

The blue line on the left of Fig.~\ref{fig:sim_turin} represents the wide 
screen mentioned in Section \ref{expturin}. We assumed that the pedestrians are 
attracted to the screen in order to have a better view of the football match. 
Thus, a (small) desire force pointing towards the screen was included at 
the beginning of the simulation. This force equaled $m\,v_d/\tau$ for 
the standing still individuals ($v(0)=0$), according to Eq.~(\ref{desired}). We 
further assumed that the pedestrians were in a relaxed state at the beginning 
of 
the simulation, and therefore, we set $v_d=0.5\,$m/s \cite{Dorso1}. This value 
accomplished a local density that did not exceed the maximum expected for 
outdoor events, say, 3-4~people/m$^2$.\\

The pedestrian in black in the middle of the crowd in Fig.~\ref{fig:sim_turin} 
represents the \textit{fake bomber} appearing in the video. He is responsible 
for triggering the panic contagion at the beginning of the simulations. For 
simplicity, we assumed that he remained still during the panic spreading 
process. \\

The pedestrians in red in Fig.~\ref{fig:sim_turin} are responsible for 
shouting the alert, as they are very close to the \textit{fake bomber} (less 
than $1\,$m). They were initially set to the panic state in the simulation (see 
Section \ref{relaxed_anxious} for details). \\ 

Recall that the event takes place outdoor. Piazza San Carlo, however, is 
surrounded by walls (as can be seen in Fig.~\ref{fig:original_frame_turin}). 
We considered along the simulations that the crowd always remained inside the 
\textit{piazza} and no other pedestrian were allowed to get inside during the 
process. \\

\subsubsection*{\label{sim_virginia}The Charlottesville scenario}

The initial conditions for simulating the incident at Charlottesville are 
somewhat different from those detailed in Section \ref{sim_turin}. The 
pedestrians are now placed at random positions within certain limits around the 
street crossing (see Fig.~\ref{fig:sim_virginia}). But, in order to 
counterbalance the social repulsion between pedestrians, a (small) inbound 
desire force was set. That is, the pedestrian's desire force pointed to the 
center of the crossing.\\ 

The total number of pedestrians appearing in Fig.~\ref{fig:sim_virginia} 
is 600. This number was computed taking into account the total number of 
occupied cells and the amount of pedestrians per cell of the first frame of the 
video (see caption of Fig.~\ref{fig:pixel_frame} for details). Due to the low 
image quality of Fig.~\ref{fig:original_frame}, we were not able to distinguish 
if more than two pedestrians per cell. Thus, we simply assumed that each red 
cell was occupied by only two pedestrians.\\

After setting the pedestrian's desire force to $v_d=0.5\,$m/s, we allowed them 
to move freely towards the center of the street crossing. This instance 
continued until a similar profile to the one in Fig.~\ref{fig:original_frame} 
was attained. \\

Then, we assumed, according to the video, that the pedestrians tried to stay at 
a fix position. Thus, we set the desired velocity to zero and allowed the 
system 
to reach the balance state before initiating the simulation. Notice that this 
condition differed slightly from the Turin condition, where the desired 
velocity 
was set to $0.5\,$m/s.  \\

The ``source of panic'' for the Charlottesville's incident corresponds to the 
\textit{offending driver} moving along the vertical direction in 
Fig.~\ref{fig:sim_virginia}. We modeled the \textit{offending driver} as a 
packed group of 21 spheres, (roughly) emulating the contour of a car (see 
Fig.~\ref{fig:sim_virginia}). The mean mass of the packed group was set to 
2000~kg. The car moved from bottom to top at 3~m/s until it reached the center 
of the street intersection. When this occurred, it stopped and remained fix 
until the end of the simulation. \\

As in the Turin situation, we assumed that those pedestrians very close to 
the car (that is, less than 1~m) entered into panic immediately, and thus, they 
were initially set to the panic state. When the car stopped, the ``source of 
panic'' was switched off.  \\

The streets are considered as open boundary conditions. This means that the 
pedestrians are able to rush away from the crossing as far as they could. \\

\subsection{\label{relaxed_anxious} The simulation process}

The videos that capture the panic spreading over the crowd let us classify the 
pedestrians into those moving relaxed or those moving anxiously. These are 
qualitative categories that can be easily recognized through the pedestrian's 
behavioral patterns. An accurate value for the inner stress $M$ seems not to 
be possible from the videos. Thus, we assume that the pedestrians may be in one 
of two possible states: relaxed or in panic. The former means that his (her) 
desired velocity does not exceed a fix threshold  $v_d^\mathrm{lim}$, or 
$M_\mathrm{lim}$, according to Eq.~(\ref{inner_strength}). The latter means 
that the individual surpassed this threshold.   \\

We already mentioned in Section \ref{sim_turin} that the desired velocity  
$v_d=0.5\,$m/s is in correspondence with either accepted literature values for 
relaxed individuals and the expected local density for approximately 900 
individuals.  Hence, we set $v_d^\mathrm{lim}=0.5\,$m/s as a reasonable 
limit for the pedestrian to be considered relaxed. This limit is supposed to be 
valid for either the Turin and the Charlottesville incident, since the total 
number of pedestrians involved in each event and the expected maximum local 
density are similar in both cases. \\

\begin{figure}
\includegraphics[width=\columnwidth]{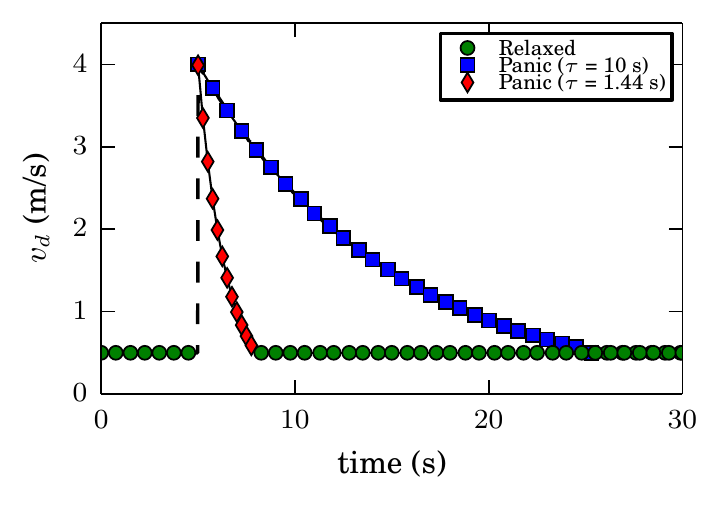}
\caption{\label{plotvd} Time evolution of the desired velocity $v_d$ for a 
single individual. The green color corresponds to the relaxed state, while the 
red color corresponds to the panic state. At $t=5\,$s the $v_d$ switches from 
$0.5\,$m/s to $4\,$m/s due to panic contagion. No further stimulus is received 
after the contagion. The desired velocity decays exponentially from $t>5\,$s 
resembling the stress decay. The red diamond and blue square symbols represents 
two exponential characteristic times: $\tau$~=~1.44~s and $\tau$~=~10~s, 
respectively. At $t\simeq 10\,$s and $t\simeq 25\,$s (depending the value of 
$\tau$) the desired velocity attains $v_d^\mathrm{lim}=0.5\,$m/s and the 
individual returns to the relaxed state. }
\end{figure}

For simplicity, $v_d^\mathrm{min}$ and $v_d^\mathrm{max}$ 
(Eq.~(\ref{inner_strength})) were set to zero and $4\,$m/s, respectively, in 
all the simulations. The maximum velocity $v_d^\mathrm{max}=4\,$m/s corresponds 
to reasonable anxiety situations appearing in the literature 
\cite{Helbing4,Dorso3,Dorso4,Dorso6}. \\

Fig.~\ref{plotvd} illustrates the time evolution for the desired velocity 
$v_d(t)$ of an individual who switches from the relaxed state to the 
panic state (see caption for details). Notice that the increase in the inner 
stress is implemented as an (almost) instantaneous change in his (her) 
desired velocity due to panic contagion. This corresponds to the 
contagion process. Once in panic, however, the stress decay phenomenon 
applies, regardless of any other neighbor expressing fear. The stress 
decay stops when the individual settles to the relaxed state, that is, 
when the $v_d$ returns back to $0.5\,$m/s. When this occurs, the pedestrian 
moves randomly at this desired velocity until the end of the simulation. \\

In order to determine the experimental value of the characteristic time $\tau$, 
we measured the time required by an anxious pedestrian to recovers his (her) 
relaxed state ($t_c$). According the analysis of the videos, 
in the Turin's case anxious pedestrians returns to a relaxed state after 20 
seconds. \\

But, in the Charlottesville case, near pedestrians (less than 5~m) from the car 
relaxed after 3 seconds. However, far away pedestrians arrive to the relaxed 
state after 20 seconds. Recall from Section \ref{expvirginia}, that near 
individuals to the car has more information about the incident nature than far 
away ones. Thus, when the car stops, near individuals recovers his relaxed 
state unlike the far away pedestrians that continues in panic. \\

We computed the experimental value of the characteristic time $\tau$ for each 
scenario using Eq.~(\ref{tau}) with $v_d^\mathrm{min}$~=~0~m/s, 
$v_d^\mathrm{max}$~=~4~m/s and $v_d(t_c)$~=~0.5~m/s. In the Turin's case, 
$\tau$ equals, approximately, to 10~seconds, while in the Charlottesville's 
case the characteristic time equals to 1.44~seconds and 10~seconds for near and 
far away pedestrians, respectively. \\

Notice that an anxious pedestrian has more or less ``influence'' over his 
neighbors according his (her) information level about the incident. That is, 
for example an individual near to the car can spread his (her) fear emotion 
over a relaxed pedestrian during, only, 3 seconds. In other words, during the 
time that were in panic ($t_c$). \\


\subsubsection*{The panic contagion process}
\label{sec:process}

The panic contagion process was implemented as follows. First, we associated an 
effective contagion stress $\mathcal{P}^{(i)}$ to each relaxed individual, 
according to Eq.~(\ref{eqn_contagion_prob}). That is, we computed the fraction 
of neighbors in the panic state $k$ to the total number of neighbors $n$ within 
a fix contagion radius of $2\,$m (from the center of mass of the corresponding 
relaxed pedestrian $i$). Second, we randomly switched the relaxed pedestrians 
to the panic state, according to the associated effective contagion 
stress $\mathcal{P}^{(i)}$. The $\mathcal{P}^{(i)}$ values were updated at each 
time step (say, $0.05\,$s). \\  
 
Notice that this contagion process may be envisaged as a 
Susceptible-Infected-Susceptible (SIS) process. The Susceptible-to-Infected  
transit corresponds to the (immediate) increase of $v_d$ from $0.5\,$m/s to 
$4\,$m/s (with effective contagion stress $\mathcal{P}^{(i)}$). The 
Infected-to-Susceptible transit corresponds to the stress decay from $4\,$m/s 
back to $0.5\,$m/s. \\

We want to remark the fact that the emotions received by an individual in the 
panic state were neglected, and thus, did not affect the stress decay process. 
This should be considered a first order approach to the panic contagion 
process. \\

\subsubsection*{Simulation software}
\label{sec:lammps}

The simulations were implemented on the {\sc Lammps} molecular dynamics 
simulator \cite{plimpton}.  {\sc Lammps} was set to run on multiple processors. 
The chosen time integration scheme was the velocity Verlet algorithm with a 
time step of $10^{-4}\,$s. Any other parameter was the same as in previous 
works (see Refs.~\cite{Dorso4,Dorso3}). \\

We implemented special modules in C++ for upgrading the {\sc Lammps} 
capabilities  to attain the ``social force model'' simulations. We simulated 
between 60 and 90 processes for each situation (see figures caption for 
details). Also, the processes lasted between 10~s and 20~s according each 
analysis. Data was recorded at time intervals of 0.05~s. The recorded 
magnitudes were the pedestrian's positions and their emotional state (relaxed 
or anxious) for each evacuation process. \\

\section{\label{results}Results}

This section exhibits the results obtained from either real life situations and 
computer simulations. Two sections enclose these results in order to discuss
them in the right context. We first analyze the Turin case (Section 
\ref{Turin_results}), while the more complex one (Charlottesville, Virginia) is 
left to Section \ref{Virginia_results}. \\  

\subsection{\label{Turin_results} Turin}

\subsubsection{\label{strength_parameter}The contagion stress parameter}

\begin{figure}
\includegraphics[width=\columnwidth]{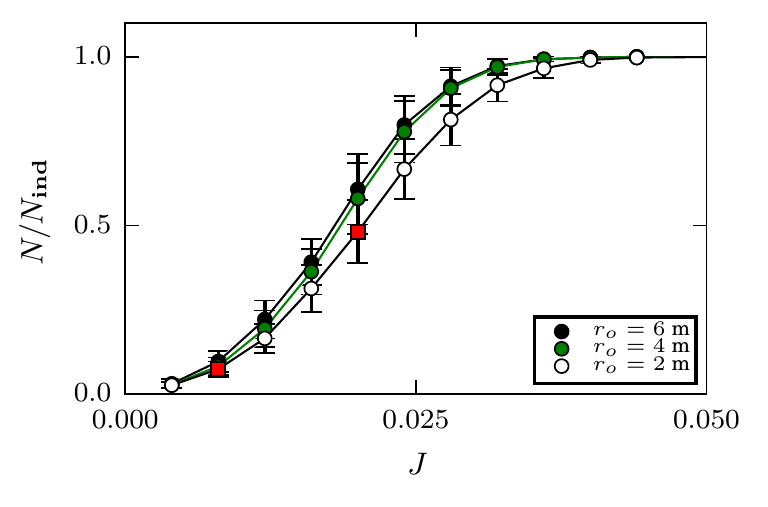}
\caption{\label{fig:anxious_turin} Normalized number of anxious pedestrians 
during the first 20~s of the escaping process as a function of the contagion 
stress $J$ for $r_o$~=~2~m, 4~m and 6~m. $N$ is the number of anxious 
pedestrians. The plot is normalized with respect to the total number of 
individuals ($N_{ind}$~=~925). $J$-values of 0.01 and 0.02 are indicated in red 
color (and squared symbols). Mean values were computed from 60 realizations. 
The 
error bars corresponds to $\pm\sigma$ (one standard deviation).}
\end{figure}

As a first step, we measured the mean number of anxious pedestrians during 
the first $20\,$s of the escaping process for a wide range of contagion 
stresses ($J$). This is shown in Fig.~\ref{fig:anxious_turin}. As can be seen, 
the number of anxious pedestrians increases for increasing contagion stresses. 
 That is, as pedestrians become more susceptible to the fear emotions from his 
(her) neighbors, panic is allowed to spreads easily among the crowd. \\

The fraction of pedestrians that switch to the anxious state exhibits three 
qualitative categories as shown in Fig.~\ref{fig:anxious_turin}. For $J$ 
ranging between 0 to 0.01, no significant spreading appears. But this scenario 
changes rapidly for the $J$ (intermediate) range between 0.01 and 0.03. The 
slope in  Fig.~\ref{fig:anxious_turin} experiences a maximum throughout this 
interval. However, if the stress becomes stronger (say, above 0.03), the 
majority enters into panic regardless of the precise value of $J$. A seemingly 
threshold for this is around $J=0.04$. \\

Notice that Fig.~\ref{fig:anxious_turin} is in agreement with the experimental 
Turin value for the mean contagion stress ($J=0.100\pm0.055$, see Section 
\ref{experimental}). The panic situation at Piazza San Carlo, as observed from 
the videos, shows that all the pedestrians moved to the panic state. The 
snapshot in Fig.~\ref{fig:media_frame_turin} illustrates the situation a 
while after the (fake) bomber called for attention.  \\ 

The panic contagion shown in Fig.~\ref{fig:anxious_turin} does not appear to 
change significantly for increasing contagion radii. We explored situations 
enclosing only first neighbors ($2\,$m) to situations enclosing as far as 
$6\,$m. The number of pedestrians in panic always attained a maximum slope at 
almost the same $J$ value for all the investigated situations. This value 
(close to 0.025) seems to be an upper limit for any weak panic spreading 
situation, or the lower limit for any widely spreading situation. We may 
hypothesize that two qualitative regimes may occur for the panic propagation in 
the crowd.  \\

Following the above working hypothesis, we turned to study any morphological 
evidence for both regimes in Section~\ref{turin_morphology} . \\

\subsubsection{\label{turin_morphology}The escaping morphology}

Our next step was to examine the anxious pedestrian's spatial distribution for 
the Piazza San Carlo scenario. The corresponding videos show that the 
individuals tried to escape radially from the (fake) bomber (see 
Fig.~\ref{fig:media_frame_turin}). Thus, the polar space binning (\textit{i.e.} 
cake slices) centered at the (fake) bomber seemed the most suitable framework 
for inspecting the crowd morphology piece-by-piece. We binned the 
\textit{piazza} into $N_\mathrm{bins}=30$ equally spaced pieces as shown in 
Fig.~\ref{fig:sim_turin}. The angle between consecutive bins was 
$\varphi=360^{\circ}$/$N_\mathrm{bins}=12^{\circ}$. \\

\begin{figure*}
\subfloat[\label{fig:digramas_turin}]{
\includegraphics[width=\columnwidth]{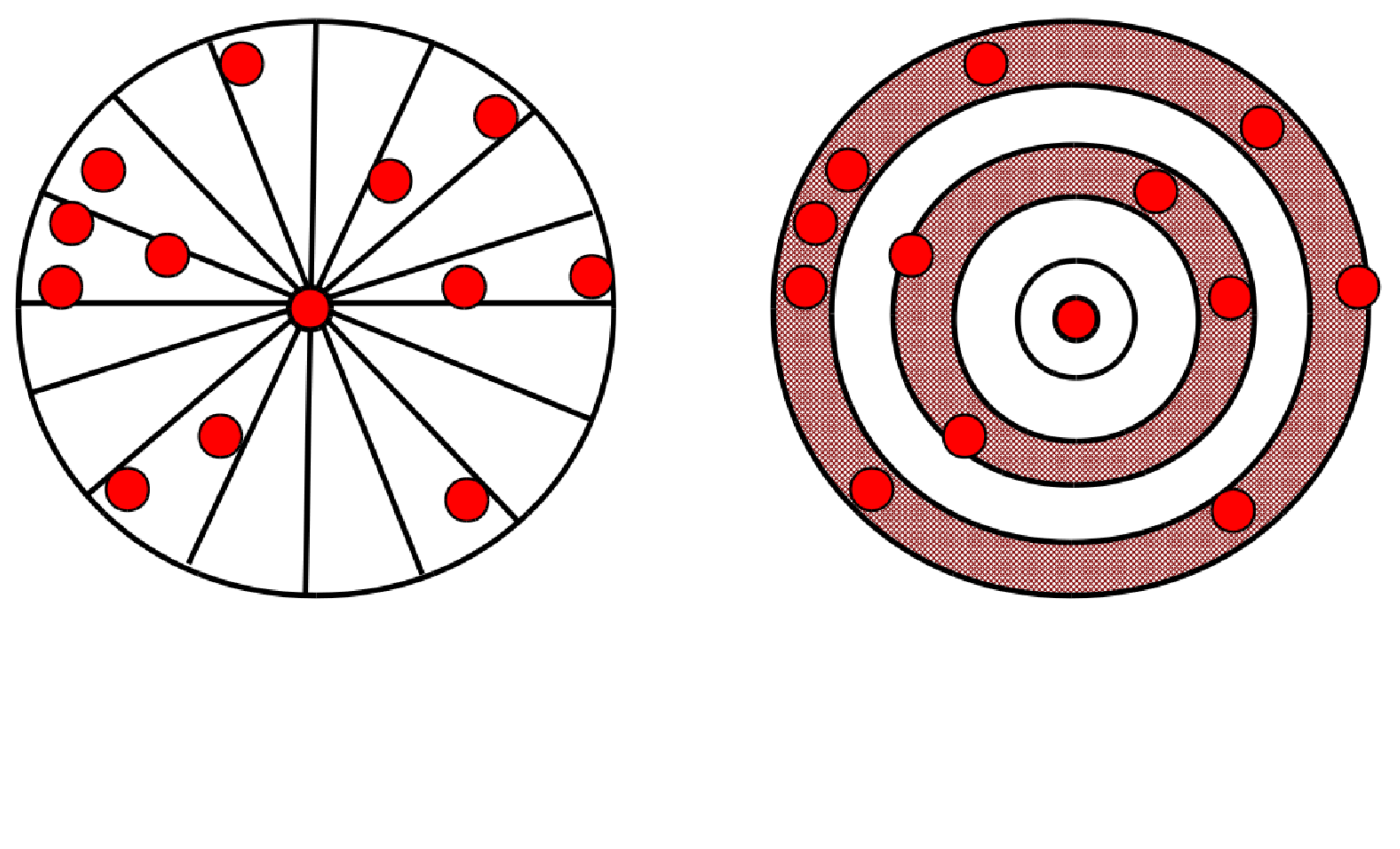}
}
\hfill
\subfloat[\label{fig:rays}]{
\includegraphics[width=\columnwidth]{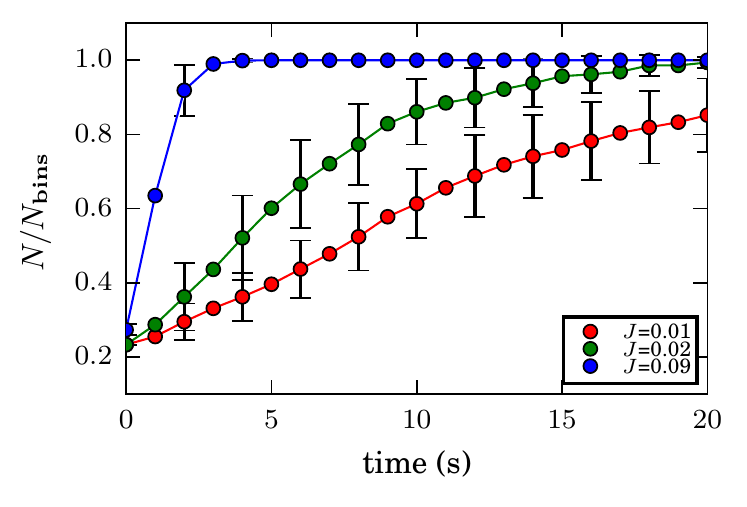}
}
\caption{\label{fig:rays_turin} (a) Schematic representation of the radial 
(left) and circular (right) bins (see text for more details). The red circles 
represent the position of many anxious pedestrians. The fake bomber is place at 
the center of the region. (b) Fraction of occupied radial bins by anxious 
pedestrians vs. time (in seconds) for $J=0.01$, 0.02 and 0.09. $N$ is the 
number of occupied radial bins (see text for more details). The plot is 
normalized with respect to the total number of radial bins 
($N_\mathrm{bins}=30$). Mean values were computed from 60 realizations. The 
error bars corresponds to $\pm\sigma$ (one standard deviation).} 
\end{figure*}

Fig.~\ref{fig:rays_turin} exhibits the number of occupied bins or slices 
(normalized by the total number of bins) occupied \textit{at least} by one 
anxious pedestrian. Three different contagion situations are represented there. 
These situations attain the qualitative categories mentioned in 
Section~\ref{strength_parameter}. That is, $J=0.01$ for low panic spreading, 
$J=0.02$ for an intermediate spreading and $J=0.09$ for wide panic spreading 
(see caption of Fig.~\ref{fig:rays_turin} for details).  \\

According to Fig.~\ref{fig:rays_turin}, the number of occupied bins (slices) 
increases monotonically during the escaping process. This means that panic 
propagates in all directions (from the bomber) until nearly all the 
slices become occupied.  However, the slopes for each situation are quite 
different. As the contagion stress $J$ increases, the bins become occupied 
earlier in time (higher slopes). For the most widely spread situation 
($J=0.09$) 
all the slices become occupied before the first 5 seconds, meaning that we may 
expect escaping pedestrians in any direction during most of the contagion 
process. \\  

\begin{figure*}
\subfloat[$J=0.01$\label{fig:snapshots_j_001_b}]{
\includegraphics[scale=0.25]{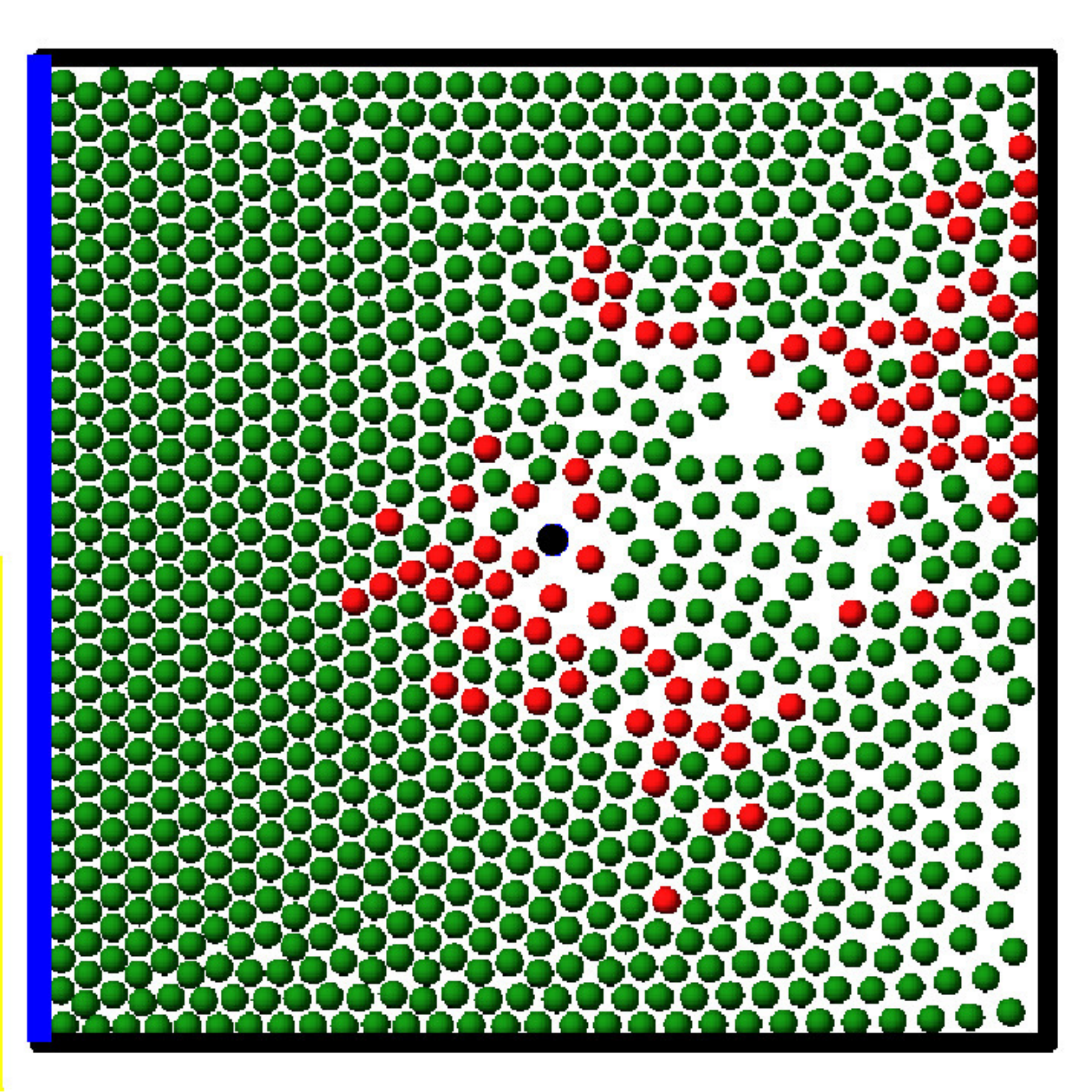}
}
\hfill
\subfloat[$J=0.02$\label{fig:snapshots_j_002_b}]{
\includegraphics[scale=0.25]{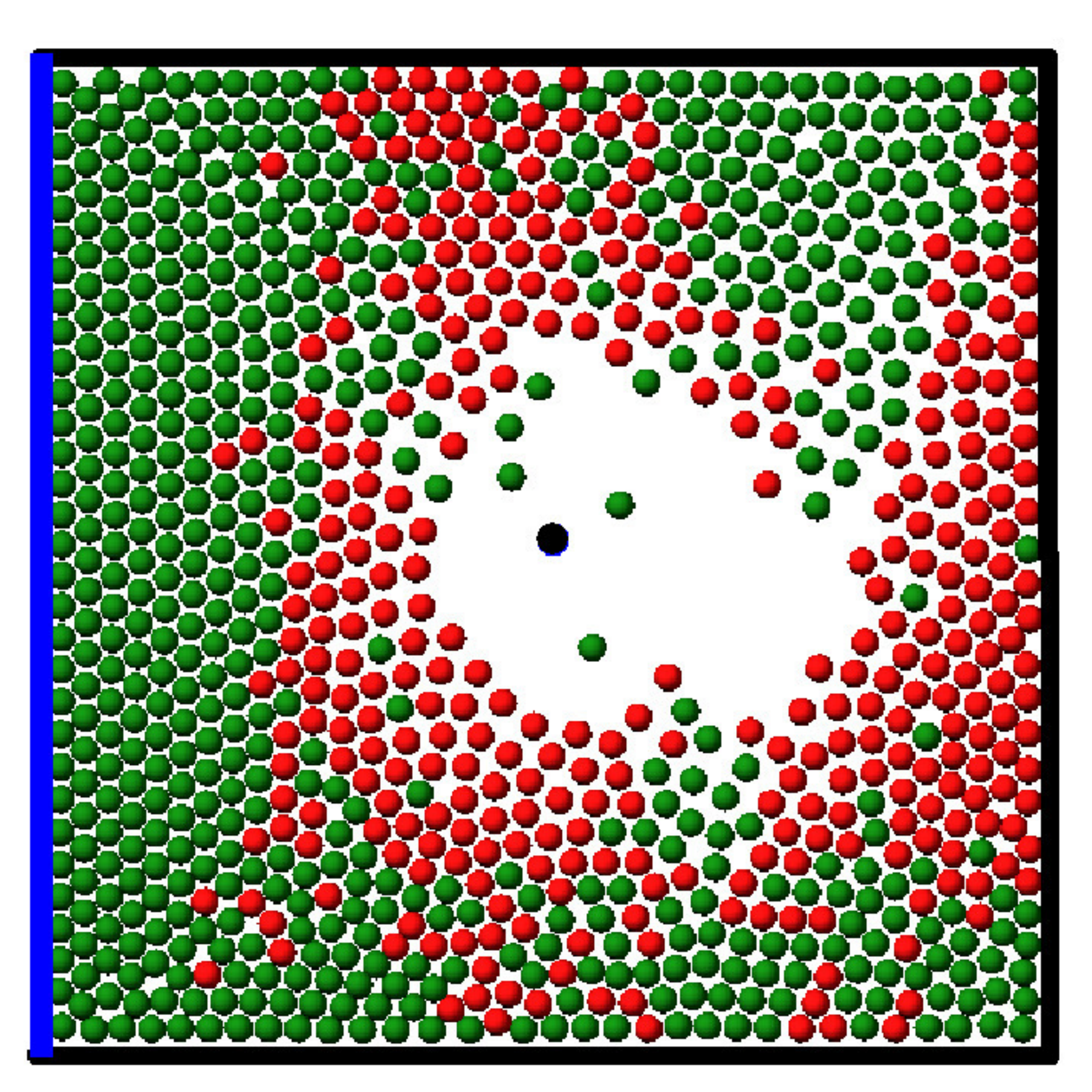}
}
\hfill
\subfloat[$J=0.09$\label{fig:snapshots_j_009_b}]{
\includegraphics[scale=0.25]{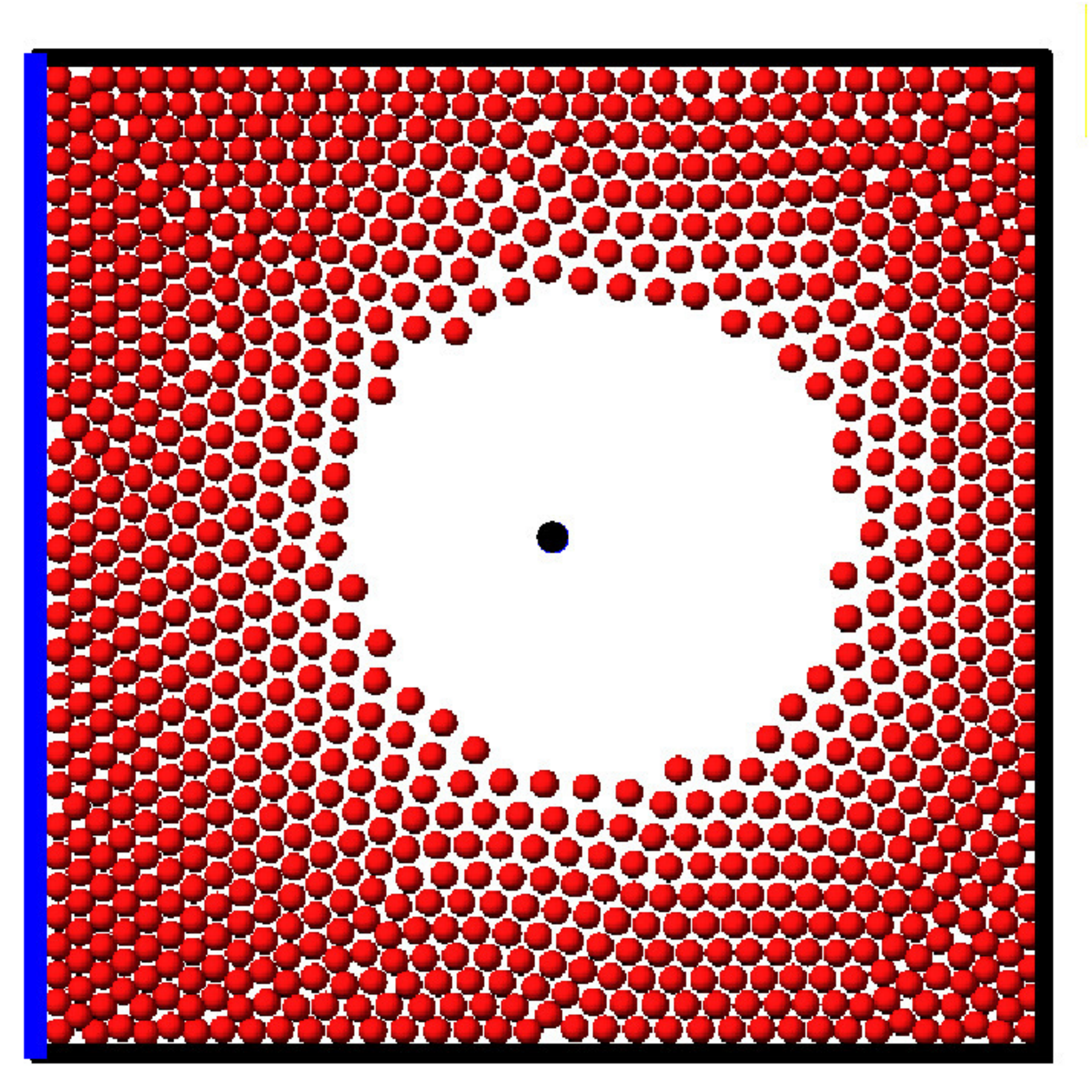}
}
\caption{\label{fig:types_evacuation} Snapshots of different escaping processes 
for three values of contagion stress in the first 15 seconds. The different 
colors of the circles represents the anxiety state of each pedestrian. 
Relaxed and panic pedestrians are represents in green and red circles, 
respectively. The \textit{fake bomber} is place at the center of the region and 
is represented in black circle. Relaxed pedestrians desire to reach the screen 
located on the left (blue line).} 
\end{figure*}

Fig.~\ref{fig:types_evacuation} represents the aforementioned three situations 
after $15\,$s since the (fake) bomber shout (see caption for details). These 
snapshots may be easily compared with the corresponding slice occupancy plot 
exhibited in Fig.~\ref{fig:rays_turin}.  \\

Fig.~\ref{fig:snapshots_j_001_b} corresponds to the lowest contagion stress 
($J=0.01$). We can see a somewhat ``branching'' pattern for those pedestrians 
in panic (red circles). That is, a branch-like configuration is present around 
the (fake) bomber. From the inspection of the whole process through an 
animation, we further noticed that these branches could be classified into two 
types (see below). The ``branching'' profile is also present in 
Fig.~\ref{fig:snapshots_j_002_b} for $J=0.02$, although this category exhibits 
an extended number of pedestrians in panic. The highest contagion stress 
category ($J=0.09$), instead, adopts a circular profile (see 
Fig.~\ref{fig:snapshots_j_009_b}). \\

The ``branching'' profile observed for $J=0.01$ and $J=0.02$ may be associated 
to the positive slopes in Fig.~\ref{fig:rays_turin}. Likewise, the circular 
profile for $J=0.09$ can be associated to the flat (blue) pattern therein. This 
suggests, once more, that two qualitative regimes may occur for the panic 
propagation in the crowd, as hypothesized in Section~\ref{strength_parameter}. 
Low contagion stresses correspond to the (qualitative) branch-like regime, 
while high contagion stress correspond to the (qualitative) circular-like 
regime. The snapshot in Fig.~\ref{fig:media_frame_turin} clearly shows a 
circular-like regime, as expected for the obtained experimental value of $J$. \\

The branching-like profile in Piazza San Carlo is not completely symmetric 
since the pedestrian's density is higher near the screen area (on the left 
of Fig.~\ref{fig:experimentaldataTurin} and Fig.~\ref{fig:types_evacuation}) 
than in the opposite area. The pedestrians near the screen can not move away as 
easily as those in the opposite direction. Thus, the panic contagion near the 
screen occurs among almost static pedestrians, while the contagion on the 
opposite area occurs among moving pedestrians. Both situations, although 
similar in nature, produce an asymmetric branching. We labeled as 
\textit{passive} branching the one near the screen, and \textit{active} 
branching the one in the opposite direction. \\  

It may be argued that since the $J=0.09$ pattern in Fig.~\ref{fig:rays_turin} 
exhibits a positive slope at the very beginning of the contagion process and a 
vanishing slope a few seconds after (say, $5\,$s), the association of 
branch-like to low $J$, and circular-like to high $J$, is somehow artificial. 
This is not true, as explained below. \\

We further binned the \textit{piazza} into circular sectors around the (fake) 
bomber as shown in Fig.~\ref{fig:rays_turin} (see caption for details). We 
carried out a similar analysis as in Fig.~\ref{fig:rays_turin}, but for the 
sectors. Say, we computed the number of occupied sectors at each time-step. 
The results were similar as for the slices (not shown). This means that both 
(slices and sectors) behavioral patterns are strongly correlated (for any fixed 
$J$). \\

The number of occupied sectors, somehow, indicates the speed of the radial 
propagation. Thus, the circular-like profiles correspond to higher speeds than 
the branch-like profiles, and consequently, it is not possible to associate low 
stress to circular profiles (or high stress to branch profiles). 
The propagation velocity will simply not match. Indeed, the circular profile 
appears only at high contagion stresses (for this \textit{piazza} geometry). \\

We may summarize the investigation so far as follows. The panic spreading 
dynamic may experience important (qualitative) changes according to the 
``efficiency'' of the alerting process between neighboring pedestrians. This is 
expressed by the contagion stress parameter $J$. The Piazza San Carlo video, 
and our simulations, show that panic propagates weakly for low values of $J$. 
This produces a branch-like, slow panic spreading around the source of danger 
(for a simple geometry). However, if  $J$ exceeds (approximately) 0.025 the 
panic contagion spreads freely in a circular-like profile (for a simple 
geometry). The propagation also becomes faster. \\

It should be emphasized that $J\sim 0.025$ is an approximate threshold, but 
well formed circular-like profiles appear, in our simulations, for stresses 
above 0.03. Stresses beyond 0.04 exhibit similar profiles as those for $J\sim 
0.04$. These results are valid for contagion radii between $2\,$m and $6\,$m. \\

Recall that the increase in the ``inner stress'' is the underneath mechanism 
 allowing the panic to spread among the crowd. The ``emotional decay'', 
however, 
seems not to play a relevant role in Piazza San Carlo (and in our simulations). 
This is because the experimental characteristic time for the ``emotional 
decay'' 
is $\tau=10\,$s, allowing anxious pedestrians to settle back to the relaxed 
state after $20\,$s (see Fig.~\ref{plotvd}). \\

We will discuss in Section~\ref{Virginia_results} a geometrically complex 
situation where either the ``inner stress'' and the ``emotional decay'' plays 
a relevant role. \\

\subsection{\label{Virginia_results} Charlottesville, Virginia}

\subsubsection{\label{density_map}Density contour}

\begin{figure}
\includegraphics[width=\columnwidth]{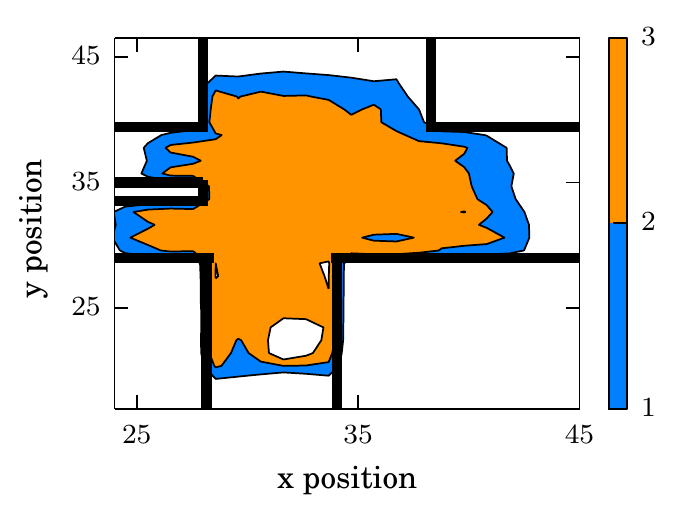}
\caption{\label{fig:density_map_virginia_sim}Mean density contour lines 
computed from 90 processes. This density map corresponds to the initial state 
of the Charlottesville simulation. The scale bar on the right is expressed in 
people/m$^2$. The black lines represent the walls. The white circular shaped 
pattern at the bottom of the image corresponds to the car location. The contour 
lines were computed on a squared grid of $1.5\,\mathrm{m}\times1.5\,\mathrm{m}$ 
and then splined to get smooth curves. Level 
colors can be seen in the on-line version only.}
\end{figure}

We first computed the discretized density pattern at the beginning of the 
simulation process in order to compare it with the video pattern shown 
in Fig.~\ref{fig:pixel_frame}. We used the same cell size as in 
Fig.~\ref{fig:pixel_frame} ($1.5\,\mathrm{m}\times1.5\,\mathrm{m}$). The 
corresponding contour density map can be seen in 
Fig.~\ref{fig:density_map_virginia_sim}. \\ 

Fig.~\ref{fig:density_map_virginia_sim} and Fig.~\ref{fig:pixel_frame} exhibit 
the same qualitative profiles. Also, the pedestrian occupancy per cell is 
similar on both figures. Notice that the middle of the region is occupied by 
two 
or more pedestrians per cell. The boundary cells, though, are occupied by a 
single pedestrian per cell in both figures. So, we may conclude that our 
initial 
configuration is qualitative and quantitative similar to the one in the video. 
\\

\subsubsection{\label{strength_parameter_Virginia}The contagion stress 
parameter $J$}

\begin{figure*}
\subfloat[\label{fig:anxious_virginia_low_J}]{
\includegraphics[width=\columnwidth]{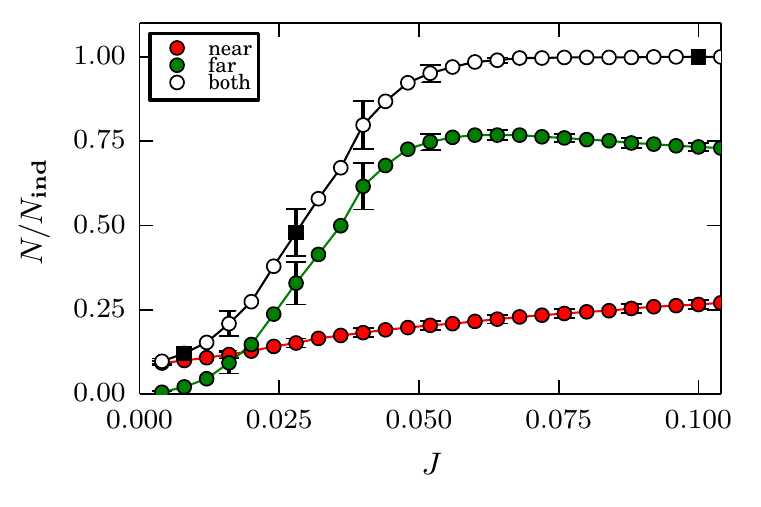}
}
\hfill
\subfloat[\label{fig:anxious_virginia_high_J}]{
\includegraphics[width=\columnwidth]{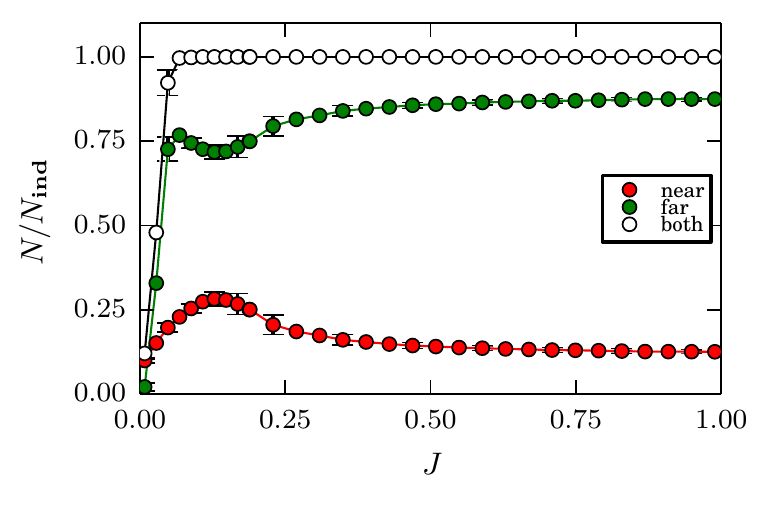}
}
\caption{\label{fig:anxious_virginia} Normalized number of anxious 
pedestrians during the first 15~s of the escaping process, as a function of the 
contagion stress $J$. $N$ is the number of anxious pedestrians. The plot is 
normalized with respect to the total number of individuals ($N_{ind}$~=~600). 
The red color corresponds to panicking pedestrians close to the car (less than 
5~m), while the green color corresponds to those far away from the car (more 
than 5~m). The white symbols correspond to the (normalized) set of all 
individuals in panic. Results for $J$~=~0.010, 0.028, and 0.1 are indicated 
in black color (and squared symbols). Mean values were computed from 60 
realizations. The error bars corresponds to $\pm\sigma$ (one standard 
deviation).} 
\end{figure*}

Our next step was, as in the Turin case, to compute the mean number of 
anxious pedestrians during the first 15 seconds of the escaping processes as 
a function of the contagion stress ($J$). The results can be seen in 
Fig.~\ref{fig:anxious_virginia}.  \\

We observe that, likewise the Turin case, that the total number of anxious 
pedestrians (hollow symbols and black line) increases for increasing contagion 
stresses. From the comparison between Fig.~\ref{fig:anxious_turin} and 
Fig.~\ref{fig:anxious_virginia_low_J} we may realize that both situations 
exhibit the same qualitative patterns for the total number of anxious 
pedestrians. However, the slope for the Charlottesville situation 
is somewhat lower with respect to the Turin situation (see 
Fig.~\ref{fig:anxious_turin}). \\

In order to explain this slope discrepancy we computed, separately, the (mean) 
number of anxious pedestrians close to the car (\textit{i.e.} source of panic) 
and those far away from the car. Recall from Section \ref{simulations} that the 
former correspond to better informed pedestrians than the latter. The 
computation of the number of ``near'' anxious pedestrians actually include 
those pedestrians that get into panic very close to the car (less than 
1~m). \\

We may recognize from Figs.~\ref{fig:anxious_virginia_low_J} and 
\ref{fig:anxious_virginia_high_J} the same three qualitative categories 
mentioned in Section \ref{strength_parameter}, according to the contagion 
stress value. Notice, however, that the anxious pedestrians now settle 
to the relaxed state after 3 seconds (if near the car) or 20 seconds (if far 
away from the car). We will examine these regimes in the following sections. \\

\subsubsection*{The low contagion stress regime}
\label{sec:low}

For $J$ ranging between 0 to 0.02, most of the pedestrians that get anxious are 
close to the car, while the far away pedestrians remain in a relaxed state. 
This means that panic does not spread homogeneously over all the crowd. \\

Recall from Section~\ref{sim_virginia} that individuals located very close to 
the car (less than 1~m) get into panic immediately. So, as the car moves across 
the crowd, the panic propagates first over these nearby individuals. This 
explains why, in Fig.~\ref{fig:anxious_virginia_low_J}, there is a small number 
of anxious pedestrians for extremely low contagion stresses ($J$~$\sim$~0).\\

Notice that this small group of anxious pedestrians represent the first source 
of panic inside the crowd (regardless of the car). As the susceptibility to 
fear emotions increase, their neighbors get into panic. But, due to their 
rapid fear decay (3 seconds), their influence on the surrounding neighbors is 
low. This is the reason for the smooth increment of the near anxious 
pedestrians. \\

Besides, we can observe from Fig.~\ref{fig:anxious_virginia_low_J} that 
the number of far away pedestrians getting into panic is not significant. 
Any pedestrian located far away from the car may only get anxious if panic 
surpasses his (her) contagion radius. So, if the number of ``near'' anxious 
pedestrians is low while also relaxing quickly (\textit{i.e.} 3 seconds), then 
the ``probability'' that panic reaches far away pedestrians from the car is 
indeed very low. This explains the low number of far away pedestrians that get 
anxious during this interval (less than 0.02). \\

\subsubsection*{The intermediate contagion stress regime}
\label{sec:intermediate}

The panic spreading scenario changes if $J$ ranges between 0.02 and 0.05. 
Along this interval, the total number of anxious pedestrians (white circles) 
increases abruptly. We can observe that this corresponds essentially to the 
increase in the amount of far away anxious pedestrians. Indeed, the 
number of near anxious pedestrians shown in 
Fig.~\ref{fig:anxious_virginia_low_J} exhibits a smooth increment that cannot 
explain the abrupt increase of the total number of anxious individuals. \\

Notice that an increment in the number of anxious ``far away'' pedestrians 
becomes possible (at high contagion stresses) due to the significant time 
window that they spend surrounded by other ``far away'' anxious pedestrians 
(say, 20 seconds). Thus, the compound effect of high susceptibility to fear 
emotions and the long lasting time decays ($t_c$) explains the sharp increase 
in the number of anxious pedestrians. \\

\subsubsection*{The high contagion stress regime}
\label{sec:high}

Finally, if the contagion stress becomes intense (say, above 0.05), most of 
the individuals get into panic regardless of the precise value of $J$. Thus, as 
in the Turin situation, we may consider a seemingly threshold for this 
regime around $J$~=~0.07. Fig.~\ref{fig:anxious_virginia_high_J} shows, 
however, that two noticeable behaviors appear whether the contagion stress is 
(roughly) below $J$~=~0.2 or not (despite the fact that the majority enters 
into 
panic).\\

Below $J$~=~0.2, the number of pedestrians that get into panic near the 
car increases for increasing contagion stresses, while above this 
threshold the corresponding slope in Fig.~\ref{fig:anxious_virginia_high_J} 
changes sign. The number of far away anxious pedestrians exhibit, though, a 
small ``U'' shape and a positive slope for $J\gg0.2$ (see 
Fig.~\ref{fig:anxious_virginia_high_J}). \\

The increase in the number of individuals that get into panic near the car just 
below the threshold $J$~=~0.2 attains for the increase in the susceptibility 
to fear emotions.  But, above $J$~=~0.2, the situation is somewhat 
different. The contagion stress is so intense that panic propagates rapidly 
into de crowd. People standing as far as 5~m from the car may switch to an 
anxious state, and thus, they get into panic \textit{before} the car 
(\textit{i.e.} the source of panic) approaches them. Our simulation movies (not 
shown) confirm this phenomenon. We further realized that many of the 
anxious individuals located near the car and computed into the red curve in 
Fig.~\ref{fig:anxious_virginia_high_J} at $J\leq0.2$ may actually move to the  
green curve at extremely intense stresses $J\gg0.2$. \\

The above research may be summarize as follows. We identified three scenarios 
according to the contagion stress. If the susceptibility to fear emotions is 
low (below 0.02), the panic spreads over a small group of pedestrians located 
very close to the car. In the case of an intermediate contagion stress ($J$ 
between 0.02 and 0.05), the number of pedestrians that get into panic far away 
from the car increases abruptly. Above $J$~=~0.05, the panic spreads over all 
the crowd. \\

The propagation velocity of the fear among the crowd is related to the 
contagion stress ($J$). As the susceptibility to fear emotions increases, 
the panic spreading velocity also increases. So, if pedestrians are very 
susceptible to fear emotions, just a small number of individuals is capable of 
spreading panic over the whole crowd. \\

\subsubsection{\label{virginia_morphology}The escaping morphology}

\begin{figure*}[!htbp]
\subfloat[Area.\label{fig:area_Minkowski}]{
\includegraphics[width=1.0\columnwidth]
{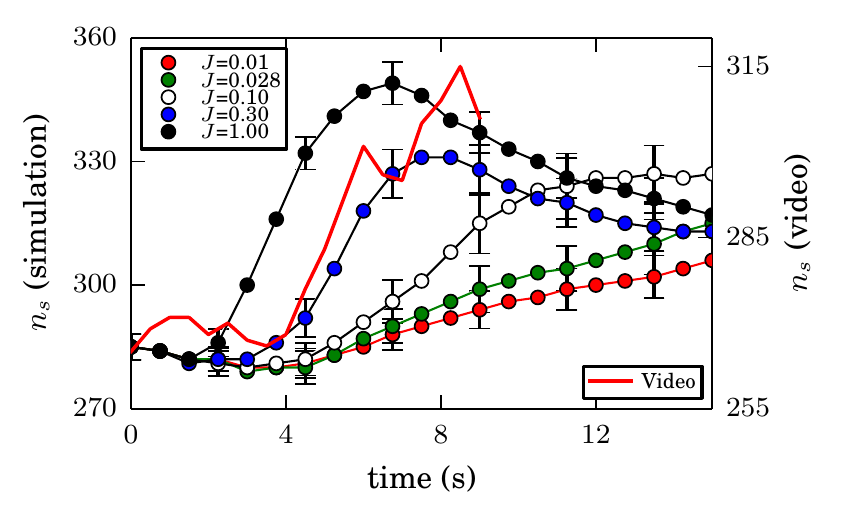}
}
\subfloat[Perimeter. \label{fig:perimeter_Minkowski}]{
\includegraphics[width=1.0\columnwidth]{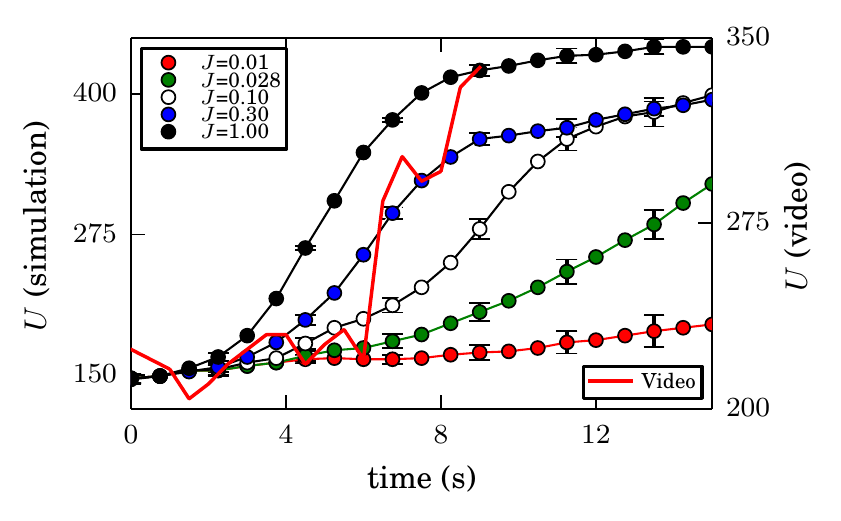}
}
\caption{\label{fig:Minkowski_functionals} Minkowski functionals corresponding 
to the (a) area ($n_s$) and (b) perimeter ($U$) (see Section~\ref{Minkowski} 
for more details). The red, green, white blue and black circles corresponds to 
the Charlottesville simulation. The red line corresponds to the experimental 
data extracted from the video incident. In both cases, the cell 
area was 1.5~m$^2$. The duration of the Charlottesville video was 10 seconds. 
Mean values were computed from 90 realizations. The error bars corresponds to 
$\pm\sigma$ (one standard deviation).} 
\end{figure*}

In Section \ref{strength_parameter_Virginia} we computed the total number of 
anxious pedestrians as a function of the contagion stress $J$. Now, we 
examine the pedestrian's spatial distribution. We computed the Minkowski 
functionals (area and perimeter) for different contagion stresses. The 
results are shown in Fig.~\ref{fig:Minkowski_functionals}. \\

The examined situations attain the same qualitative categories mentioned in 
Section \ref{strength_parameter_Virginia}. That is, $J$~=~0.01 for low panic 
spreading and $J$~=~0.028 for an intermediate spreading, and the two cases 
($J$~=~0.1 and $J$~=~0.3) for the highly intense situation. We also analyzed 
the 
limiting case ($J$~=~1). No distinction was made at this point between relaxed 
or anxious pedestrians. \\

Recall from Section~\ref{Minkowski}, that the area is the number of occupied 
cells by, at least, one pedestrian. Fig.~\ref{fig:area_Minkowski} shows two 
qualitatively different patterns, one before the first 4 seconds and the other 
one after this time period. The former exhibits a slightly negative slope, 
while a positive slope can be seen in the latter (at least for a short time 
period). 
\\

The first 4 seconds in the contagion process correspond to the time period 
since the car strikes against the crowd until it stops. So, we may associate 
the decrease in the area with the movement of the pedestrians  next to the 
car. The process animations actually show that these individuals group 
themselves as the car moves towards the crowd. \\

The slope changes sign after the first 4 seconds, meaning an increase of the 
occupied area (see Fig.~\ref{fig:area_Minkowski}). This corresponds, according 
to our animations (not shown), to pedestrians running away from each other. 
The greater the contagion stress, the sharper the slope. Since these slopes 
represent somehow the escaping velocity, Fig.~\ref{fig:area_Minkowski} 
expresses the fact that people try to escape faster as they become more 
susceptible to fear emotions (at least during this short time period). \\

Fig.~\ref{fig:perimeter_Minkowski} exhibits the results for the computed
perimeter. This functional informs us on the length of the (supposed) boundary 
enclosing the crowd. Unlike the area, the perimeter appears as an increasing 
function of time (for the inspected values of $J$). Furthermore, as the 
susceptibility to fear emotions increases, the faster the perimeter 
widens.  \\

The real life data included in Fig.~\ref{fig:Minkowski_functionals} matches 
(qualitatively) the simulated patterns. Indeed, simulations corresponding to 
high contagion stresses appear to match better. Specifically, the Minkowski 
functionals computed for $J$~=~0.30 exhibit the best matching patterns. 
Notice, however, that the scales of the experimental data and our simulations 
 are different (see Fig.~\ref{fig:Minkowski_functionals}). This scale 
discrepancy is entirely due to the differences in the size of the occupancy 
cells corresponding to experimental data and to our simulations. \\

\begin{figure*}
\subfloat[$J=0.01$\label{fig:snapshots_j_001_virg}]{
\includegraphics[scale=0.25]{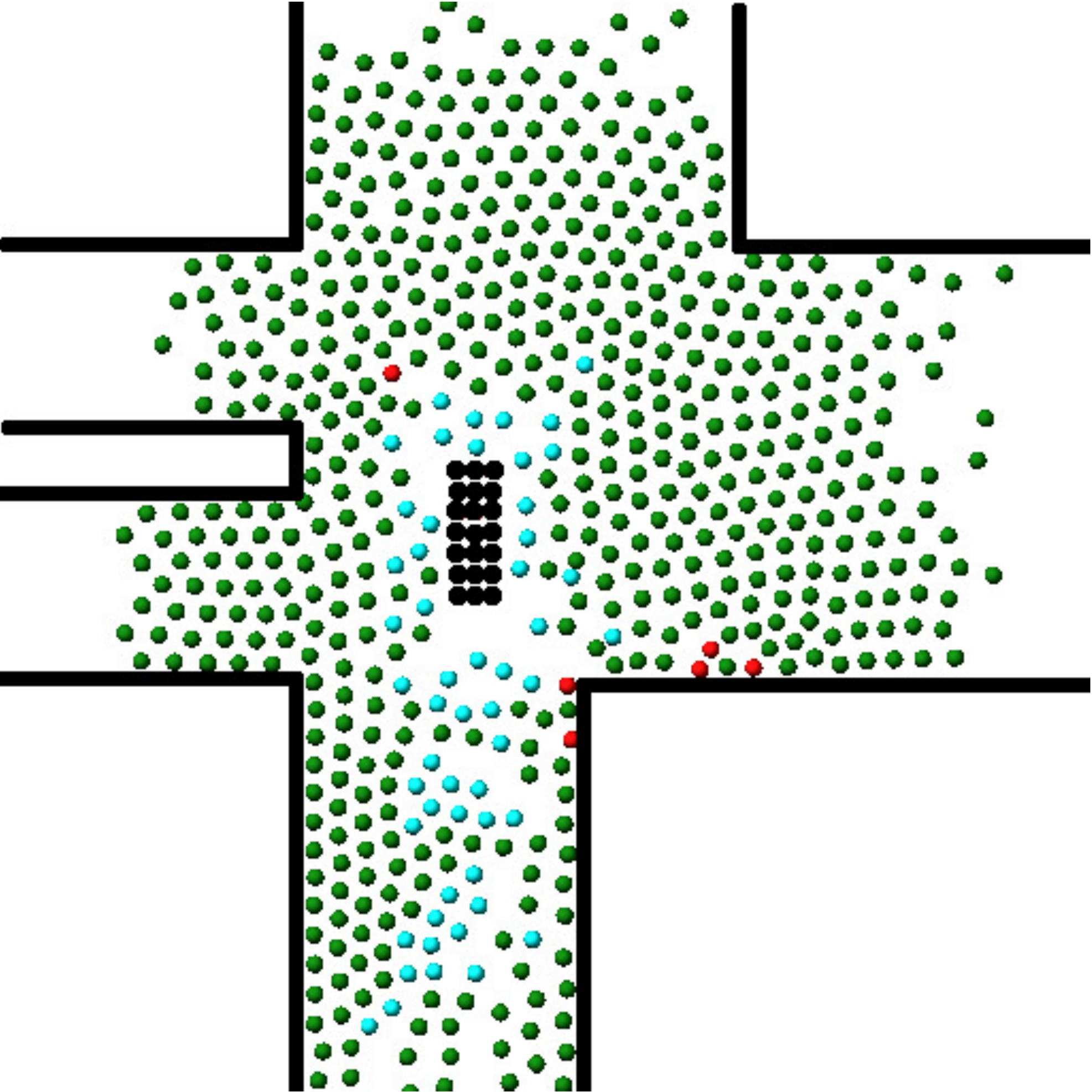}
}
\hfill
\subfloat[$J=0.028$\label{fig:snapshots_j_0028_virg}]{
\includegraphics[scale=0.25]{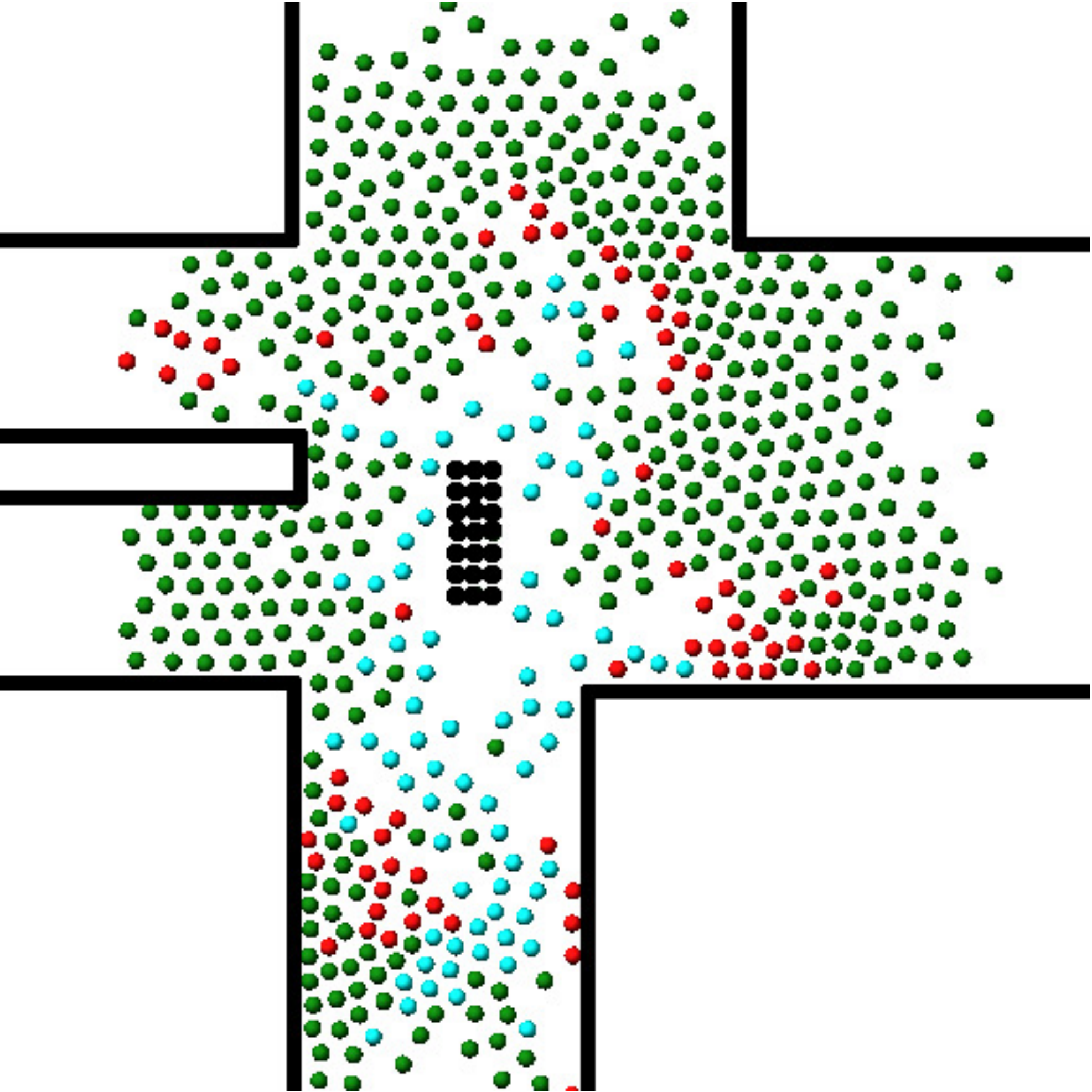}
}
\hfill
\subfloat[$J=0.3$\label{fig:snapshots_j_030_virg}]{
\includegraphics[scale=0.25]{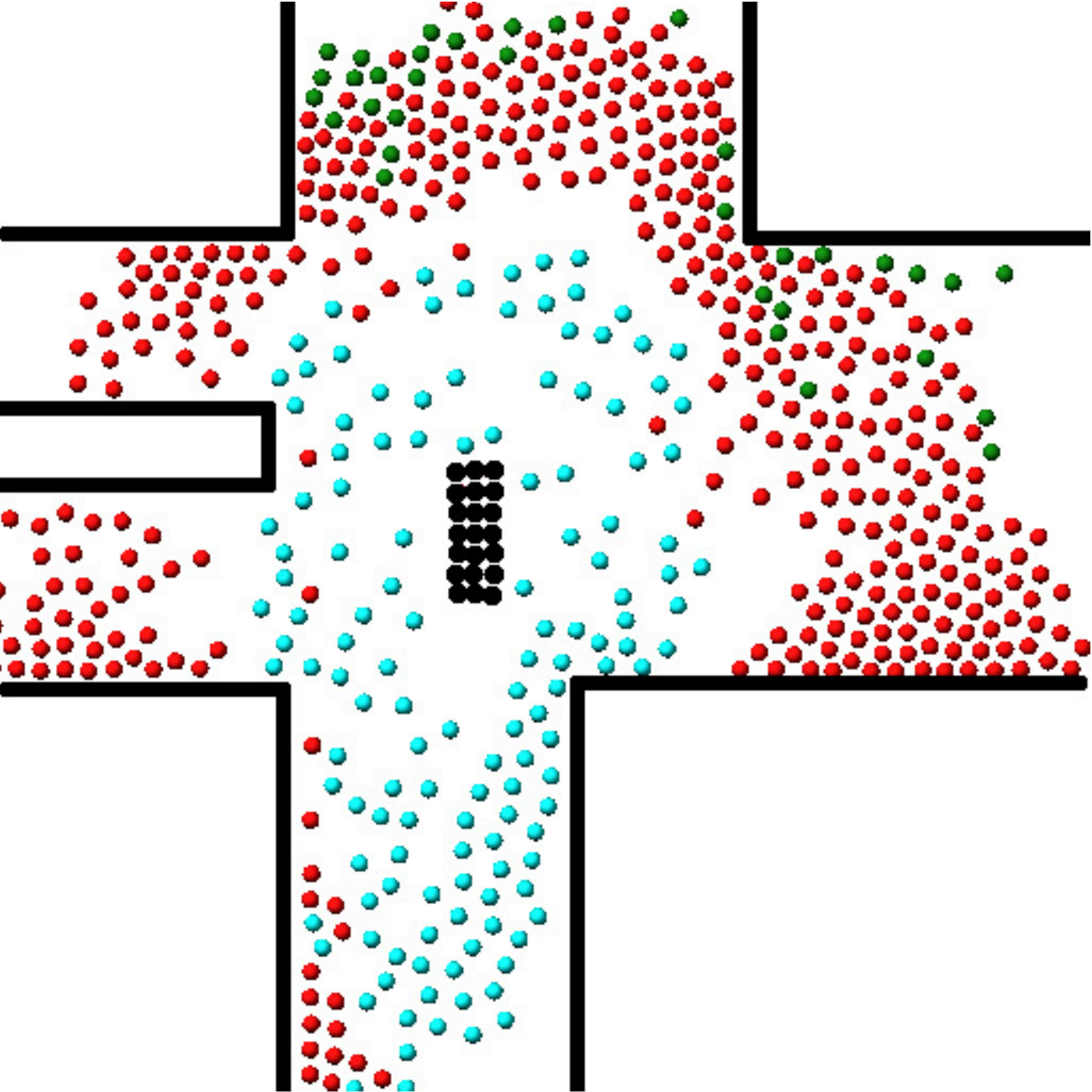}
}
\caption{\label{fig:snapshots_virginia} Snapshots of different escaping 
processes for low ($J$~=~0.01), intermediate ($J$~=~0.028) and high 
($J$~=~0.3) contagion stress in the first 10 seconds. The 
different colors of the circles represents the anxiety state of each 
pedestrian. Relaxed and panic pedestrians are represents in green and red 
circles, respectively. The cyan circles represents the recovered pedestrians. 
That is, individuals that in the past were in panic but now are relaxed due the 
stress decay. The \textit{offending driver} is represented by black circles. 
The solid lines represents the walls and the row of cars located on the middle 
left of the image (see Fig.~\ref{fig:experimentaldataCharlottesville}).} 
\end{figure*}

We finally examined the process animations for low ($J$~=~0.01), intermediate 
($J$~=~0.028) and high ($J$~=~0.3) contagion stresses separately. These values 
correspond to the symbols in black color in Fig.~\ref{fig:anxious_virginia}. 
The complementary snapshots are shown in Fig.~\ref{fig:snapshots_virginia}, 
captured after 10 seconds from the beginning of the process. We chose this 
time interval in order to differentiate the three situations more easily (see 
caption for details). \\

Fig.~\ref{fig:snapshots_j_001_virg} corresponds to the lowest contagion 
stress ($J$~=~0.01). As already shown from 
Fig.~\ref{fig:anxious_virginia_low_J}, only a small number of pedestrians 
gets into panic (due the low susceptibility to fear emotions). These 
pedestrians are colored in cyan in Fig.~\ref{fig:snapshots_j_001_virg} (on-line 
version only), and correspond to people standing close to the car path. No 
dramatic differences appear between the profiles shown in 
Fig.~\ref{fig:sim_virginia} and Fig.~\ref{fig:snapshots_j_001_virg}. Thus, we 
may expect a smooth slope for the Minkowski functionals 
(see Fig.~\ref{fig:Minkowski_functionals}). \\

Fig.~\ref{fig:snapshots_j_0028_virg} shows a somewhat different scenario 
due to $J$~=~0.028 (the intermediate contagion stress). We realize that an 
increasing number of pedestrians are now in panic. Many pedestrians that 
appeared as relaxed in Fig.~\ref{fig:snapshots_j_001_virg}, have now become 
anxious because of the fear emotions from the individuals located near the car. 
However, the occupied area did not change significantly from 
Fig.~\ref{fig:snapshots_j_001_virg}. The perimeter, instead, is expected to 
increase because of the voids left back by the panicking pedestrians inside the 
crowd (see the red circles in the on-line version of 
Fig.~\ref{fig:snapshots_j_0028_virg}).  \\

The more stressing scenario in shown in Fig.~\ref{fig:snapshots_j_030_virg}. 
The whole crowd gets into panic for $J=0.3$. This situation is comparable to 
the Turin incident (above $J$~=~0.09), despite the obvious geometrical 
differences. Thus, qualitatively speaking, Fig.~\ref{tracks} and 
Fig.~\ref{fig:snapshots_j_030_virg} show ``similar'' crowd profiles. \\

A few conclusions can be outlined from the above analysis. As in the Turin 
situation, we found different scenarios according the pedestrian's 
susceptibility to fear emotions. For low contagion stresses, the panic spreads 
over a small group of pedestrians standing close to the source of panic (the 
car). As the contagion stress ($J$) increases, the influence of these nearby  
individuals on their neighbors become more relevant. Thus, more pedestrians get 
into panic. If $J$ is above 0.05, the fear spreads over all the crowd. \\

Despite the fact that above $J$~=~0.05 all the pedestrians becomes in panic, 
there are two qualitatively different regimes, bounded by a threshold at 
(roughly) $J$~=~0.2. We found that below $J$~=~0.2, the role of the 
pedestrians near the source of panic (say, the better informed pedestrians) 
is a relevant one, since they are the mean for propagating panic deep inside de 
crowd. But, above $J$~=~0.2, the contagion stress is so intense that panic 
propagates rapidly into the crowd even though a minimum number of individuals 
near the car get into panic.  \\

\section{\label{conclusions}Conclusions}

The contagion of panic in a crowd is usually thought to propagate like a 
disease among a social group. But reliable parameters for properly testing 
this hypothesis are not currently available. This investigation introduced two 
real panic-contagion events, in order to arrive to a trusty model for the panic 
propagation. Our work was carried out in the context of the ``social force 
model''.   \\

The contagion of panic offered a challenge to the emotional mechanism operating 
on the pedestrians. We only included the ``inner stress'' and ``stress decay'' 
as the main processes triggered during a panic situation. Although the 
simplicity of this model, we attained fairly good agreement with the real 
panic-contagion events. \\      

We handled the coupling mechanism between individuals through the contagion 
stress parameter $J$. This parameter appears to be responsible for increasing 
the ``inner stress'' of the individuals. Our first achievement was getting a 
real (experimental) value for $J$. The value for the Piazza San Carlo event was 
$0.1\pm 0.055$. \\

We further noticed through computer simulations that $J$ controls the 
contagion dynamics. The Piazza San Carlo event illustrates the dynamic arising 
for high values of $J$, where everyone moves away from the source of stress. 
However, this might not be the case for low values of $J$. Only a small number 
of pedestrians will escape from danger, although many will roughly stay at 
their current position. The whole impression will be like random ``branches'' 
(the pedestrians in panic) moving away from the source of danger. We actually 
concluded that $J\sim 0.03$ is roughly the limit between both dynamics. \\ 

Our simulations attained qualitatively correct profiles for the escaping crowd 
either at Piazza San Carlo and the Charlottesville street crossing. But these 
profiles are geometry-dependent, and therefore, not a unique profile could be 
established for any value of $J$ at different incidents. We know (for now) that 
geometries similar to Piazza San Carlo  may produce branch-like profiles 
($J<0.03$) or circular-like profiles ($J>0.03$). \\ 

The ``stress decay'' depends on the nature of the source of panic (say, whether 
if it corresponds to a fake alert or not) and the amount of information that 
the pedestrians get from this source. That is, far away pedestrians from the 
igniting point of panic (fake bomber in the Turin situation, or the offending 
driver in the Charlottesville situation) may not have enough information on the 
nature of the incident, but nearby pedestrians may get a more precise picture 
of the incident. The cleared this picture becomes to them, the faster they are 
allowed to settle down, and thus, the shorter the characteristic decay time. \\

We realized, however, that a shorter characteristic time actually prevents the 
panic from spreading. This was not the case at the Piazza San Carlo, since the 
fake bomber was not (directly) at the sight of the pedestrians (who where 
watching the football match). The Charlottesville incident, however, exhibited 
two groups of individuals, according to the available information. We noticed 
that the group near the source of danger attained a shorter $\tau$ than the 
others, preventing this group from escaping. \\



What we learned from the street crossing incident at Charlottesville is that 
the resulting pedestrian's dynamic is a consequence of the competing effects 
of the ``inner stress'' (increased by contagion stimuli) and the ``stress 
decay''. Both are essential issues for a trusty contagion model. The parameters 
$J$ and $\tau$ appear as the most relevant ones within our model. \\

The $J$ and $\tau$ parameters may not always be available because of poor 
recordings or missing data. We experienced this difficulty with the video of 
the Charlottesville incident. But the experimental geometrical functionals, 
like the area or the perimeter, allowed the estimate of $J$ by comparison with 
respect to simulated data ($J\sim0.3$).   \\   

We want to remark that different contagion radii (between $2\,$m and $6\,$m) 
did not produce significant changes on our simulations. This was unexpected, 
and thus, we may speculate that ``spontaneous'' contagion out of the usual 
contagion range may not produce dramatic changes, if the probability of 
``spontaneous'' contagion is small. \\

\begin{acknowledgments}
This work was supported by the National Scientific and Technical 
Research Council (spanish: Consejo Nacional de Investigaciones Cient\'\i ficas 
y T\'ecnicas - CONICET, Argentina) grant number PIP 2015-2017 GI, founding 
D4247(12-22-2016). 
\end{acknowledgments}

\appendix

\section{\label{appendix_1}The contagion efficiency}

Any individual among the crowd may increase his (her) anxiety level if his 
(her) neighbors are in panic. This is actually the propagation mechanism for 
panic: one or more pedestrians express their fear, alerting the others of 
imminent danger. The latter may get into panic and thus, a ``probability'' 
exists for getting into panic.   \\

We hypothesize that the ``probability to danger'' (\textit{contagion 
efficiency}) is the cumulative effect of the alerting neighbors. That is, if 
$k$ 
pedestrians among $n$ neighbors are expressing fear, then the contagion 
efficiency $\mathcal{P}_n$ of an individual is 

\begin{equation}
 \mathcal{P}_n=p_n(1)+p_n(2)+...+p_n(n)
\end{equation}

\noindent where $p_n(k)$ represents the contagion efficiency of $k=1,2,...,n$ 
pedestrians (among $n$ neighbors) expressing fear. The distribution for 
$p_n(k)$ 
is a Binomial-like distribution if any neighbor expresses panic 
with fixed contagion efficiency $p$, regardless of the feelings of other 
neighbors. If the feelings of any neighbor (among $n$ pedestrians) is not 
completely independent of the other neighbors, $p_n(k)$ should be assessed as a 
Hypergepmetric-like distribution. \\

For the purpose of simplicity we assume that the Binomial-like distribution is 
a valid approximation for the $p_n(k)$ computation. Consequently, 
 
\begin{equation}
 \mathcal{P}_n=\displaystyle\sum_{k=1}^n\left(\begin{array}{c}
                                         n \\
                                         k \\
                                        \end{array}\right)
p^k(1-p)^{n-k}=1-(1-p)^n\label{eq:p_n}
\end{equation}

The mean value of neighbors expressing fear $\langle k\rangle$ is $np$. Thus, 

\begin{equation}
 \mathcal{P}_n=1-\bigg(1-\displaystyle\frac{\langle 
k\rangle}{n}\bigg)^n\label{exponential}
\end{equation}

It is worth noting that this expression holds for a fix value of $n$. That is, 
the contagion efficiency $\mathcal{P}_n$ is conditional to the amount of 
neighboring individuals $n$. The contagion efficiency for any number of 
neighbors $n=1,2,...,M$ is \\

\begin{equation} 
\mathcal{P}=\displaystyle\sum_{n=1}^M\mathcal{P}_n\,\pi_n\label{conditional}
\end{equation}

\noindent where $\pi_n$ means the contagion efficiency that there are $n$ 
neighbors surrounding the anxious pedestrian. Notice that the expression 
(\ref{conditional}) neither includes the term for $n=0$, nor the terms above 
$M$. The situation $n=0$ is not considered here since it corresponds to a
``spontaneous'' contagion to danger. The situation $n>M$ corresponds to far 
away individuals, and thus, not really capable of alerting of danger. The 
limiting value $M$, however, is supposed to be related to a pertaining distance 
and the the crowd packing density.\\     

There is no available information on the values of $\pi_n$, although it may be 
written as the ratio $\pi_n=z_n/M$ (number of current neighbors with respect 
to the maximum number of neighbors). \\

Recalling Eq.~(\ref{exponential}), the contagion efficiency $\mathcal{P}_n$ may 
be expanded as 

\begin{equation}
 \mathcal{P}_n=1-(1-np+...+p^n)=p\,f_n(p)
\end{equation}

The function $f_n(p)$ stands for the summation 

\begin{equation}
f_n(p)=n-\displaystyle\frac{n(n-1)}{2}\,p+...+p^{n-1}\label{fn}
\end{equation}

Each contributing terms in $f_n(p)$ may be envisage as the alert to danger 
due to groups of individuals of increasing size (for a fix number of neighbors 
$n$). Notice, however, that the expression (\ref{fn}) holds if the feelings 
between neighboring pedestrians are completely independent. Otherwise, the 
function $f_n(p)$ should be considered unknown. \\

The overall contagion efficiency reads

\begin{equation}
\mathcal{P}=\displaystyle\sum_{n=1}^M
\displaystyle\frac{z_n}{M}\,\displaystyle\frac{\langle 
k\rangle}{n}\,f_n(p)\simeq 
J\,\bigg\langle\displaystyle\frac{k}{n}\bigg\rangle\label{overall}
\end{equation}

\noindent where $J$ represents an \textit{effective stress} for 
the propagation, since it expresses in some way the efficiency of the alerting 
process. That is, no panic propagation will occur for vanishing values 
of $J$, while the pedestrian susceptibility to fear emotions will become more 
likely as $J$ increases. The stress $J$ may depend, however, on the 
probability $p$. Appendix \ref{appendix_2} shows that this dependency is weak 
enough to be omitted in a first order approach.  \\

The fraction $\langle k/n\rangle$ corresponds to the mean fraction of neighbors 
expressing fear with respect to the total number of neighbors. This mean 
fraction is computed over all the possible number of neighbors, according to 
Eq.~(\ref{overall}).\\

\section{\label{appendix_2}The sampling procedure for Turin}

The effective stress $J$ may be evaluated from any real life situation. 
Details on the sampling procedure for the Turin incident at Piazza San Carlo 
are given in Section \ref{expturin}. \\

As a first step, we identified those individuals that switched to the panic 
state along the image sequence. We also identified the surrounding 
pedestrians for each anxious individual, and labeled them as neighboring 
individuals (regardless of their current anxiety state). For simplicity, we 
used the same profile (shown in Fig.~\ref{fig:analysis_frame_turin}) throughout 
the image sequence. \\

The mean fraction $\langle k/n\rangle$ was obtained straight forward from this 
data. Table \ref{table:1} exhibits the corresponding results (see second 
column). \\

\begin{table}
\caption{Data provided from the Turin video (see Section \ref{expturin} for 
details). Samples were taken at 0.5~s time intervals. The second column shows 
the number of pedestrians $n_p$ that switched to the panic state at the 
corresponding time stamp. The third column exhibits the (mean) ratio between 
neighbors in panic with respect to the surrounding neighbors. The fourth column 
corresponds to the contagion efficiency $\mathcal{P}$ computed as a 
``no-replacement'' procedure (see text). The last column corresponds to 
the \textit{contagion stress} $J$ computed from the third and fourth columns. 
The total number of individuals was $N=131$. }
\label{table:1} 
\begin{tabular}{c@{\hspace{11mm}}c@{\hspace{10mm}}c@{\hspace{10mm}}c@{\hspace{
10mm}}c}
 \hline
 $t$ & $n_p$ & $\langle k/n\rangle$ & $n_p/(N-N_p)$ & $J$ \\
 \hline 
 0.5 & 1  & 0.17 & 0.0077 & 0.0453 \\ 
 1.0 & 1  & 0.20 & 0.0077 & 0.0385 \\ 
 1.5 & 5  & 0.43 & 0.0391 & 0.0909 \\ 
 2.0 & 5  & 0.42 & 0.0406 & 0.0967 \\ 
 2.5 & 2  & 0.13 & 0.0169 & 0.1300 \\ 
 3.0 & 4  & 0.55 & 0.0345 & 0.0627 \\ 
 3.5 & 6  & 0.36 & 0.0536 & 0.1489 \\ 
 4.0 & 13 & 0.64 & 0.1226 & 0.1916 \\ 
 4.5 & 11 & 0.68 & 0.1183 & 0.1740 \\ 
 5.0 & 10 & 0.52 & 0.1219 & 0.2344 \\  
 5.5 & 22 & 0.63 & 0.3055 & 0.4849 \\  
 6.0 & 29 & 0.90 & 0.5800 & 0.6444 \\  
 6.5 & 15 & 0.88 & 0.7143 & 0.8117 \\  
 \hline
\end{tabular}
\end{table}

Notice that the surrounding pedestrians actually correspond to the most inner 
ring of pedestrians enclosing the anxious individual, but not the ones within a 
certain radius. This radius, however, can be estimated from the (mean) 
packing density of the crowd. \\

The anxious pedestrians at the border of the examined area of Piazza San 
Carlo (see \label{fig:experimentaldataTurin}) are not included in Table 
\ref{table:1} since it was not possible to identify \textit{all} of their 
surrounding pedestrians. \\

The fraction of the anxious pedestrians $n_p$ to the total number of 
individuals $N$ is a suitable estimate for the overall contagion 
efficiency $\mathcal{P}$.  However, as panic propagates, the acknowledged 
anxious pedestrians $n_p$ diminish because the number of previously relaxed 
individuals reduces inside the analyzed area. Thus, the estimate of 
$\mathcal{P}$ follows a sampling ``without replacement'' procedure. That is, 
the 
fraction estimate is $n_p/(N-N_p)$, where $N_p$ corresponds to the number of 
individuals in panic until the previous time step. \\

\begin{figure}
\includegraphics[width=1.0\columnwidth]{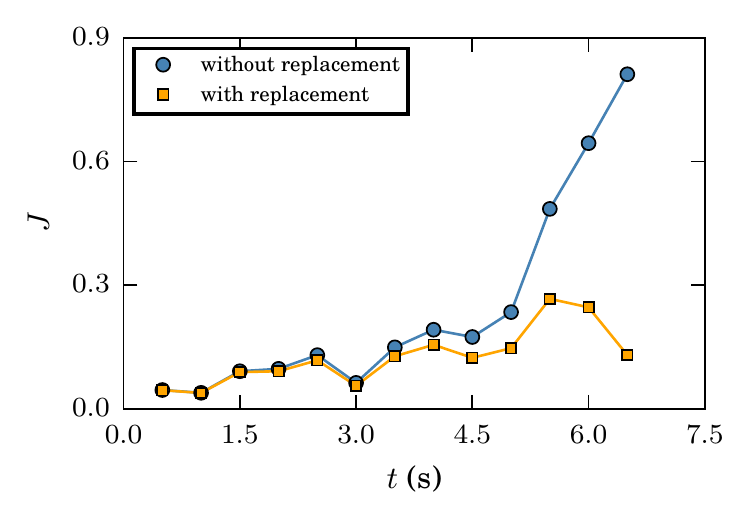}
\caption{\label{intensity} (Color on-line only) The contagion stress $J$ 
as a function of time $t$ in seconds (see text for details). The rounded 
symbols (in blue color) correspond to the $J$ values computed from a crowd of 
$N=131$ individuals and a sampling procedure ``without replacement'' (see Table
\ref{table:1} for details). The squared symbols correspond to the $J$ values 
computed from the same crowd, but following a sampling procedure ``with 
replacement''. The mean stress for $0.5\,\mathrm{s}\leq t\leq 4\,\mathrm{s}$ 
is $J=0.1\pm 0.055$. }
\end{figure}

Fig.~\ref{intensity} shows the effective stress $J$ computed as the ratio 
between $\mathcal{P}$ and $\langle k/n\rangle$. The contagion efficiency 
$\mathcal{P}$ was estimated either as $n_p/(N-N_p)$ (\textit{i.e.} without 
replacement) or $n_p/N$ (\textit{i.e.} with replacement). It can be seen that 
the sampling effects can be neglected for $t\leq 4\,$s. \\

The $J$ estimates exhibited in Fig.~\ref{intensity} are not completely 
stationary along the interval $0.5\,\mathrm{s}\leq t\leq 4\,\mathrm{s}$. 
However, the increasing slope is not relevant for a first order approach. The 
mean value for the effective stress along this interval is $J=0.1\pm 0.055$. 
\\

\newpage 
\bibliography{paper}

\end{document}